\documentclass[fleqn,usenatbib]{mnras}

\usepackage[T1]{fontenc}

\DeclareRobustCommand{\VAN}[3]{#2}
\let\VANthebibliography\thebibliography
\def\thebibliography{\DeclareRobustCommand{\VAN}[3]{##3}\VANthebibliography}

\usepackage{graphicx}	
\usepackage{amsmath}	
\usepackage{amssymb}	
\usepackage{colortbl}
\usepackage{pdflscape}
\usepackage{newtxtext,newtxmath}

\title[Radiation shielding of young discs]{Radiation shielding of protoplanetary discs in young star-forming regions}

\author[M. J. C. Wilhelm et al.]{
Martijn J. C. Wilhelm,$^{1}$\thanks{E-mail: wilhelm@strw.leidenuniv.nl}, Simon Portegies Zwart$^1$, Claude Cournoyer-Cloutier$^2$, Sean C. Lewis$^3$, \newauthor Brooke Polak$^{4,6}$, Aaron Tran$^5$, Mordecai-Mark Mac Low$^{6,5}$
\\
$^1$Leiden Observatory, Leiden University, P.O. Box 9513, NL-2300 RA, Leiden, the Netherlands \\
$^2$Department of Physics and Astronomy, McMaster University, Hamilton, ON, Canada \\
$^3$Department of Physics, Drexel University, Philadelphia, PA, USA \\
$^4$Institut f\"{u}r Theoretische Astrophysik, Zentrum f\"{u}r Astronomie, Universit\"{a}t Heidelberg, Heidelberg, Germany \\
$^5$Department of Astronomy, Columbia University, New York, NY, USA \\
$^6$Department of Astrophysics, American Museum of Natural History, New York, NY, USA
}

\date{Accepted XXX. Received YYY; in original form ZZZ}

\pubyear{2022}

\begin{document}
\label{firstpage}
\pagerange{\pageref{firstpage}--\pageref{lastpage}}
\maketitle

\begin{abstract} 
Protoplanetary discs spend their lives in the dense environment of a star forming region. While there, they can be affected by nearby stars through external photoevaporation and dynamic truncations. We present simulations that use the AMUSE framework to couple the Torch model for star cluster formation from a molecular cloud with a model for the evolution of protoplanetary discs under these two environmental processes. We compare simulations with and without extinction of photoevaporation-driving radiation. We find that the majority of discs in our simulations are considerably shielded from photoevaporation-driving radiation for at least 0.5 Myr after the formation of the first massive stars. Radiation shielding increases disc lifetimes by an order of magnitude and can let a disc retain more solid material for planet formation. The reduction in external photoevaporation leaves discs larger and more easily dynamically truncated, although external photoevaporation remains the dominant mass loss process. Finally, we find that the correlation between disc mass and projected distance to the most massive nearby star (often interpreted as a sign of external photoevaporation) can be erased by the presence of less massive stars that dominate their local radiation field. Overall, we find that the presence and dynamics of gas in embedded clusters with massive stars is important for the evolution of protoplanetary discs.
\end{abstract}




\section{Introduction}

The currently known population of more than 5000 exoplanets is incredibly diverse. The observationally prominent classes of Super Earth and Hot Jupiter planets are not found in our Solar System, and planetary system architectures such as compact resonant chains of terrestrial planets differ from the Solar System. The origin of this diversity is the subject of current research \citep[e.g.][]{vanElteren2019,Kruijssen2021,vanderMarel2021,Speedie2022}. 

A potential driver of this diversity is the environment in which a planetary system is born. Star formation typically takes place in a clustered environment \citep{Lada2003}, and planets form in protoplanetary discs around young stars. Star clusters dissolve into the Galactic field on typical time scales of 100 Myr \citep{Krumholz2019}, whereas planet formation is thought to take place in the first $\sim$1 Myr \citep[e.g.][]{Kruijer2014,Sheehan2018,Tychoniec2020}, and is at most constrained by the disc lifetime which is typically a few to ten megayears \citep{Mamajek2009,Michel2021,Pfalzner2022}. Although these numbers are uncertain, this implies that planet formation typically takes place while the host star is in a region of enhanced stellar density compared to the Galactic field.

One of the clearest examples of the environment influencing protoplanetary discs is proplyds. Discovered by \citet{Odell1993,Odell1994} in the Orion Nebula Cluster (ONC), proplyds are interpreted as protoplanetary discs which lose mass in a thermal wind due to irradiation by a nearby star\footnote{Originally simply a contraction of PROtoPLanetarY Disc, this term has since come to specifically refer to rapidly evaporating protoplanetary discs.}. This loss of mass from the disc due to external irradiation is termed external photoevaporation (EPE). The modelling of this process started soon after its discovery \citep{Johnstone1998,Storzer1999,Adams2004}, but modelling the wind was challenging due to the complex thermodynamics (and the chemistry coupled to that) in protoplanetary discs. \citet{Clarke2007}; \citet{Mitchell2010}; and \citet{Anderson2013} coupled calculations of the mass loss rate to time-dependent models of a viscous disc. Recently, \citet{Haworth2018b} developed the FUV Radiation Induced Evaporation of Discs (FRIED) grid. They computed the mass loss rate due to EPE for a grid of host star masses, radiation fields, disk radii, and disc masses. This allowed EPE to be studied for large populations of discs.

A recurring theme in EPE is that it is very efficient in depleting protoplanetary discs. Mass loss rates can be as high as $\sim$10$^{-6}$ M$_\odot$ yr$^{-1}$ \citep{Henney1998, Henney1999, Henney2002} in discs that appear to be at least 1 Myr old \citep{Beccari2017}. This would imply initial disc masses much greater than observed \citep{Tychoniec2018}, and discs that are potentially gravitationally unstable \citep{Haworth2020}. Simulations \citep[e.g.][]{Concha2021} find similarly short disc lifetimes. Attempts to reconcile the observed high mass loss rates at comparatively late age typically involve migration of discs into regions of high radiation fields and late formation of either discs or the massive star driving EPE \citep[e.g.][]{Winter2019}.

EPE can impact planet formation by shaping the protoplanetary disc. It can halt gas accretion and the migration of giant planets when it truncates the disc to where the planet forms \citep{Winter2022}, and it can also reduce the pebble flux in the disc inwards of the outer edge \citep{Sellek2020}. 

Another environmental influence is the disruption of a disc during a close encounter with another star. Such encounters typically result in the truncation of the disc beyond a certain radius \citep{Breslau2014}, but can also impact the inner disc by inducing spiral arms \citep{Pfalzner2003} and accretion bursts \citep{Pfalzner2008}. In the following text we will refer to this process as dynamic truncation. A number of suspected ongoing and recently occurred dynamic truncation events have been observed \citep{Cabrit2006,Kurtovic2018,Zapata2020,Dong2022, Huang2022}. 

Multiple studies have modelled these processes in star clusters, using the cluster's structure and dynamics to inform models of EPE and/or dynamic truncation. These works used either one or both of the processes, and various models for the star cluster. Some studies started from spherically symmetric stellar distributions, with e.g. \citet{Rosotti2014,Vincke2015,Vincke2016,Vincke2018,Concha2019a} investigating dynamic truncations, \citet{Winter2019} investigating EPE, and \citet{Concha2019b,Concha2021} investigating both. Other works reproduce the substructure of young stellar clusters with a fractal distribution, such as \citet{PortegiesZwart2016} in their dynamic truncation simulations and \citet{Nicholson2019,Parker2021a,Parker2021b} in their EPE simulations. Recently, \citet{Concha2022} and \citet{Qiao2022} used models that included the formation of a star cluster from a collapsing molecular cloud. \citet{Concha2022} included both EPE and dynamic truncation, but not stellar feedback on cluster gas, or radiative transfer. This required star formation to be halted manually, and prohibited them from taking into account extinction due to cluster gas. Still, the extended period of star formation resulted in discs being present that were young and massive compared to the average within the cluster. \citet{Qiao2022} included feedback from massive stars, but did not resolve individual stars. Instead, they grouped them in sink particles 0.45 pc in radius, and assumed a mean distance of half a cell size, 0.09 pc, between stars in a sink. This subgrid approach prohibited them from including dynamic truncation and gas and stellar dynamics near the stars that drive EPE. They found that gas within the cluster shields protoplanetary discs for $\sim$0.5 Myr after the start of star formation.

In this work we aim at modelling the formation and dynamics of individual stars, the evolution of their protoplanetary discs, and feedback effects of the stars on the gas. To this end we couple the protoplanetary disc population model developed by \citet{Concha2022,Wilhelm2022} to the Torch model for the formation of star clusters in collapsing molecular clouds \citep{Wall2019,Wall2020}. 

We run a series of simulations with and without radiation shielding due to the cluster gas. These runs allows us to characterise the importance of radiation shielding on the early evolution of protoplanetary discs. We investigate the relative importance of EPE and dynamic truncation, the lifetimes of protoplanetary discs, and the external photoevaporation of dust and its consequences for planet formation.

\section{Methods} \label{sec:methods}

\subsection{Star formation model}

We use the Torch model \citep{Wall2019,Wall2020}, which combines multiple codes using the AMUSE framework \citep{PortegiesZwart2009, Pelupessy2013}, to model the environment in which our population of protoplanetary discs forms and evolves. We briefly summarise the model here, and discuss changes made to enable this work. 

Torch models the collapse of a giant molecular cloud under self-gravity, the formation of stars, and feedback on the cloud from stellar winds, extreme ultraviolet (EUV, E$_\gamma=13.6+$ eV) and far ultraviolet (FUV, E$_\gamma=5.6$--$13.6$ eV) radiation, and supernovae. It models magneto-hydrodynamics (MHD) using the FLASH adaptive mesh (AMR) model \citep{Fryxell2000}, which is internally coupled to the radiative transfer code FERVENT \citep{Baczynski2015}. Stellar dynamics are modelled with the ph4 4th-order Hermite N-body code \citep{McMillan2012}, and stellar evolution with the SeBa parameterised stellar evolution code \citep{PortegiesZwart1996,Toonen2012}. The stars and gas are dynamically coupled using a variation of the BRIDGE method \citep{Fujii2007,PortegiesZwart2020} adapted to grid-based hydrodynamics.

Torch uses the sink particle method to form stars. When the gas density in a maximally refined grid cell satisfies a number of conditions \citep[those of][e.g. a density exceeding the Jeans density, and being in a converging flow]{Federrath2010}, a sink particle is placed in that cell. This sink particle will then accrete all mass in excess of the Jeans density within its radius (which is 2.5 minimum cell sizes). This Jeans density is $1.53\cdot 10^{-20}$ g cm$^{-3}$ for our medium-resolution runs, assuming a sound speed of $1.9\cdot 10^4$ cm s$^{-1}$, and scales with the squared inverse of the minimum cell size.

Sink particles in Torch represent regions of dense gas where multiple stars can form. Upon the formation of a sink, a queue of stellar masses is generated from the Kroupa initial mass function \citep[IMF;][]{Kroupa2001}. At every time step, we check if the sink is more massive than the first stellar mass in the queue. If it is, a star of that mass is spawned, and the mass is removed from the queue and the sink's mass reservoir. This is repeated until the sink is less massive than the next queued mass. The stars are spawned at the sink's position and with the sink's velocity, both with a random offset. The position offset vector has an isotropically distributed orientation, and its magnitude is uniformly distributed between 0 and the sink radius. The velocity offset is drawn from a normal distribution in each direction, with a mean of 0 and a standard deviation of the sound speed of the surrounding gas.

In this work, stars with masses $<$1.9 M$_\odot$ form with a protoplanetary disc of a mass $\sim$10\% of its host star (see Sec. \ref{sec:ppd_ics}). In order to conserve mass in this step, the star formation algorithm has been adapted such that a star with a disc is only formed once the sink has accreted enough mass to form both the star and the disc. The disc mass is subtracted from the sink mass along with the star's mass. 

For the purposes of gravitational dynamics, we add the disc mass, and any mass accreted from a disc onto its star in the disc model, to the stellar mass. Disc material lost through EPE and dynamic truncation is removed entirely from the simulation space. We neglect tidal effects of the extended nature of the disc on the gravitational dynamics.

In our simulations only stars more massive than 7 M$_\odot$ exert feedback, through EUV and FUV radiation, stellar winds, and potentially supernovae. They are then also the only drivers of EPE. We will refer to these stars as massive stars.

\subsection{Protoplanetary disc model}

We model the population of protoplanetary discs using the model of \citet{Concha2022,Wilhelm2022}. This model is based on VADER \citep{Krumholz2015}, which numerically integrates the $\alpha$-disc model \citep{Shakura1973,LyndenBell1974}. We briefly summarise the processes modelled and refer the reader to the aforementioned work for more detail.

Our model includes four mechanisms through which discs can lose mass. The first is EPE. For this process, we derive mass loss rates from interpolation on the FRIED grid \citep{Haworth2018b}, and remove mass from the outer disc edge inwards. The second is dynamic truncation, which we discuss further in \ref{sec:dyn_trnc}. The third is accretion onto the host star. The accretion rate is based on observations of discs in a young star-forming region \citep{Alcala2014}, with a decrease in time as described by \citet{Wilhelm2022}. The fourth is internal photoevaporation (IPE), where radiation from the host star drives a mass flow from the inner disc. We base mass loss rates and profiles on \citet{Picogna2019}, with host star mass scalings by \citet{Owen2012}, and stellar X-ray luminosities by \citet{Flaccomio2012}. The X-ray luminosity also decreases with time, as described in \citet{Wilhelm2022}.

We also model the EPE of dust as in \citet{Haworth2018a}. Dust grains up to a certain size can be entrained in the EPE wind. Within about 1 Myr \citep{Birnstiel2012}, small grains coagulate into larger particles and the dust reservoir becomes more resistant to entrainment. The model represents the disc's dust reservoir as a scalar value, initially 1\% of the gas mass. It loses mass proportionally to the gas EPE loss, but with an exponentially decaying factor that represents the dust growing resistant to EPE. This model does not resolve the dynamics of dust within the disc such as inward pebble drift. It also does not model the coagulation of dust into planetesimals or planets. This mass thus represents an upper limit on the amount of solids available for planet formation. Also because of this we refer to this quantity as the disc's `solid mass', though we still refer to its depletion as `dust EPE'.

We use a turbulent viscosity parameter of $\alpha_T=10^{-3}$ throughout this work. The grid of the disc model consists of 330 cells spaced logarithmically between 0.01 and 3000 au.

\subsection{Model coupling}

\subsubsection{Radiation field}

We aim to investigate the effects of extinction from cluster gas on the efficiency of EPE. FLASH performs radiative transfer of FUV and EUV radiation, including extinction. To obtain the FUV flux to which a disc is exposed we can sample the radiation field in the grid cell the disc is located in. We refer to this method as the {\it radiative} method (i.e. derived from radiative transfer). Alternatively, we can obtain the radiation field without extinction from the positions and FUV luminosities of the stars using the $r^{-2}$ scaling of the radiation flux. We refer to this method as the {\it geometric} method (i.e. derived from geometric consideration). We could set the effective cross section of FUV radiation in FLASH to 0 to obtain the limit without extinction, but this would also eliminate the feedback effect of FUV radiation on the gas. 

We use the FUV luminosity function of Torch, which has zero luminosity for non-massive stars to reduce computational cost. Compared to \citet{Concha2022}, whose disc model we build on, we do not include FUV radiation of stars between 1.9 and 7 M$_\odot$. More massive stars will eventually dominate the radiation field, but before their formation we likely underestimate the FUV radiation field.

We made an improvement to the FUV flux estimation in Torch. In the original implementation, the energy flow into a cell by a number of rays was converted into a flux by division with the area of a cubic cell face. This leads to an underestimate of the area exposed to radiation for off-axis rays, and a subsequent overestimate of the flux. We implemented an improved approximation of this area that reduced the relative bias in the flux estimate from $+0.5$ to $-0.1$. This is detailed in Appendix \ref{app:flux}.

\subsubsection{Dynamic truncation} \label{sec:dyn_trnc}

Dynamic truncations are implemented using an event-based approach. The stellar dynamics code is interrupted when two stars come closer than 0.02 pc. Their closest approach $r_\mathrm{enc}$ is estimated by computing the periastron distance of their two-body Kepler orbit. The radius to which each disc involved in the encounter is truncated, $R_t$, is computed as in \citet{Concha2022} (based on \citet{Breslau2014}):

\begin{equation} \label{eq:trnc}
    R_t = \frac{r_\mathrm{enc}}{3} \left(\frac{m_1}{m_2}\right)^{0.32},
\end{equation}

\noindent
where $m_1$ is the mass of the disc's host star, and $m_2$ is the mass of the other star. We then remove all disc material beyond that radius. We also set the collision radius of both stars to $0.49r_\mathrm{enc}$ for the rest of the coupling time step to prevent the same encounter from being picked up multiple times. We also remove 1\% of the truncated mass from the solid reservoir.

\section{Simulations} \label{sec:simulations}

\subsection{Initial conditions}

\subsubsection{Star forming region}

The initial conditions of the simulation consist of a spherical cloud of gas embedded in a uniform medium. The cloud has a total mass of $10^4$ M$_\odot$ and a radius of 7 pc, with a Gaussian density profile, and a uniform temperature of 30 K. It has a turbulent velocity field following a Kolmogorov spectrum, and a virial ratio $\alpha_v=\frac{2E_\textrm{kin}}{E_\textrm{pot}}=0.25$. The ambient medium has a hydrogen number density of 1.25 cm$^{-3}$ and a temperature of 8000 K. There is a uniform magnetic field along the $z$-axis with a magnitude of 3 $\mu$G.

\subsubsection{Protoplanetary discs} \label{sec:ppd_ics}

After a star with a disc forms in the simulation, the disc's mass density profile is initialised as:

\begin{equation}
    \Sigma\left(r,t=0\right) = \frac{\textrm{M}_\textrm{d,0}}{2\pi\textrm{R}_\textrm{d,0}\left(1-\textrm{e}^{-1}\right)}\frac{\exp\left(-r/\textrm{R}_\textrm{d,0}\right)}{r},
\end{equation}

\noindent
inside of $\textrm{R}_\textrm{d,0}$, and $10^{-12}$ g cm$^{-2}$ outside (required by VADER, but negligible compared to the disc). $\textrm{M}_\textrm{d,0}$ and $\textrm{R}_\textrm{d,0}$ are the initial disc mass and radius, respectively. The initial pressure profile is derived assuming the ideal gas law, with a mean molecular weight of 2.33 hydrogen nuclei, and the temperature profile:

\begin{equation}
    \textrm{T}\left(r\right) = 100 \textrm{ K} \left(\frac{\textrm{M}_*}{\textrm{M}_\odot}\right)^{1/4} \left(\frac{r}{\textrm{au}}\right)^{-1/2},
\end{equation}

\noindent
where $\textrm{M}_*$ is the mass of the host star.

The initial disc mass $\textrm{M}_\textrm{d,0}$ and radius $\textrm{R}_\textrm{d,0}$ are re-scaled from \citet{Wilhelm2022}. Their initial conditions represent the most massive protoplanetary discs that are stable against gravitational instability and are still similar in structure to observed discs. We scale these initial conditions to obtain a disc mass of 0.1 M$_\odot$ at a stellar mass of 1 M$_\odot$, as in e.g. \citet{Concha2022} and \citet{Qiao2022}, while still following the observational relation they were based on. The initial disc masses and radii as a function of stellar mass are then:

\begin{equation}
    \textrm{M}_\textrm{d,0} = 0.1 \textrm{ M}_\odot \left(\frac{\textrm{M}_*}{\textrm{M}_\odot}\right)^{0.73},
\end{equation}

\begin{equation}
    \textrm{R}_\textrm{d,0} = 117 \textrm{ au} \left(\frac{\textrm{M}_*}{\textrm{M}_\odot}\right)^{0.45}.
\end{equation}

Note that our initial disc radii are larger than those in \citet{Concha2021}, who use a similar power law index but a reference radius of 30 au.

The ratio of disc mass to stellar mass decreases with stellar mass. In \citet{Concha2021}, it was found that disc lifetime increased with host star mass, while observations appear to indicate the opposite trend \citep{Carpenter2006,Dahm2007,Ribas2015}. \citet{Wilhelm2022} showed that a decreasing disc mass ratio combined with more efficient IPE and accretion at higher stellar mass could reproduce this trend in low radiation environments. However, the regions investigated by \citet{Carpenter2006} and \citet{Dahm2007} both contain O and/or B type stars. In Section \ref{sec:lifetime} we investigate whether this trend is reproduced in our simulations of regions with massive stars and high radiation fields.

\subsection{Overview of runs} \label{sec:overview}

In order to control run-to-run variations we use the same hydrodynamical initial realisation but vary the realisation of the IMF that the sink particles use. As a result, star formation will start at an identical time and position in each run. The formation of different mass stars will lead to small differences in the evolution of the gas. The formation of high-mass stars at different times (and of different masses) will lead to larger differences due to feedback. These differences can be large enough to lead to differences in the star formation process \citep{Lewis2022}.

For a number of realisations of the IMF, we run one radiative simulation and one geometric simulation. These will diverge from one another after the first high-mass star forms. The difference in FUV radiation field leads to different EPE rates. Because the disc mass is included in the gravitational mass, this leads to differences in stellar dynamics, which grows exponentially due to the chaotic nature of gravitational dynamics. Due to the gravitational coupling between the stars and the gas these perturbations then propagate to the gas dynamics. If this impacts the formation of massive stars (even if just delaying or speeding up their formation) the two systems will diverge even more rapidly. It is unclear a priori whether these perturbations are large enough to influence star formation in such a way, but our results show that they are (see Figure \ref{fig:sfh_overview}).

All our runs use a minimum FLASH refinement level $l_\textrm{max}$ of 4. We run five pairs of simulations with identical IMF realisation at maximum refinement level 5 (medium resolution; runs m1r-m5r with the radiative method of computing discs' FUV radiation field, and m1g-m5g with the geometric method, together called the {\it main} runs; runs with the same IMF realisation but different radiation field method are called {\it paired} runs). This refinement level corresponds to a minimum cell size of 0.068 pc. In addition we run two simulations at maximum refinement levels 4 and 6 to investigate resolution effects (runs l3r and h3r, respectively), using the same IMF realisation as in two of the paired runs (m3r/m3g), and one run that starts from different hydrodynamical initial conditions (m6r; this run serves to compare cloud-to-cloud variations with those following from IMF sampling). These latter three runs use the radiative method, and we call these the {\it extra} runs. 

An overview of all runs is given in Table \ref{tab:overview}. 

\begin{table*}
    \centering
    \begin{tabular}{c|c|c|c|c|c|c|c|c|c|c|c}
    Run & Colour & $l_\textrm{max}$ & FUV field method & $t_*$ (Myr) & $t_{*,>7\textrm{M}_\odot}$ (Myr) & $M_{*,>7\textrm{M}_\odot}$ (M$_\odot$) & $t_\textrm{end}$ (Myr) & Stars & Discs & Disp. discs & Trunc.'s \\
    \hline
    m1r & \cellcolor[rgb]{0.121569,0.466667,0.705882} & 5 & Radiative & 1.520 & 1.788 & 8.14 & 2.480 & 1800 & 1717 & 7 & 796 \\
    m2r & \cellcolor[rgb]{1.,0.498039,0.054902} & 5 & Radiative & 1.520 & 1.995 & 7.90 & 2.130 & 499 & 482 & 0 & 563 \\
    m3r & \cellcolor[rgb]{0.172549,0.627451,0.172549} & 5 & Radiative & 1.520 & 1.850 & 11.3 & 2.300 & 940 & 893 & 0 & 153 \\
    m4r & \cellcolor[rgb]{0.839216,0.152941,0.156863} & 5 & Radiative & 1.520 & 1.830 & 16.8 & 2.390 & 1973 & 1888 & 4 & 120 \\
    m5r & \cellcolor[rgb]{0.580392,0.403922,0.741176} & 5 & Radiative & 1.520 & 1.796 & 20.6 & 2.110 & 340 & 326 & 0 & 106 \\
    \hline
    m1g & \cellcolor[rgb]{0.121569,0.466667,0.705882} & 5 & Geometric & 1.520 & 1.788 & 8.14 & 2.480 & 2590 & 2480 & 19 & 75 \\
    m2g & \cellcolor[rgb]{1.,0.498039,0.054902} & 5 & Geometric & 1.520 & 1.997 & 7.90 & 2.160 & 531 & 511 & 0 & 392 \\
    m3g & \cellcolor[rgb]{0.172549,0.627451,0.172549} & 5 & Geometric & 1.520 & 1.850 & 11.3 & 2.300 & 897 & 854 & 2 & 164 \\
    m4g & \cellcolor[rgb]{0.839216,0.152941,0.156863} & 5 & Geometric & 1.520 & 1.830 & 16.8 & 2.380 & 1764 & 1706 & 14 & 71 \\
    m5g & \cellcolor[rgb]{0.580392,0.403922,0.741176} & 5 & Geometric & 1.520 & 1.796 & 20.6 & 2.150 & 537 & 516 & 0 & 70 \\
    \hline
    m6r & \cellcolor[rgb]{0.54902,0.337255,0.294118} & 5 & Radiative & 1.380 & 1.708 & 19.2 & 2.290 & 1693 & 1636 & 13 & 557 \\
    l3r & \cellcolor[rgb]{0.890196,0.466667,0.760784} & 4 & Radiative & 1.452 & 1.705 & 16.8 & 2.530 & 2705 & 2591 & 7 & 172 \\
    h3r & \cellcolor[rgb]{0.498039,0.498039,0.498039} & 6 & Radiative & 1.631 & 1.893 & 11.8 & 2.080 & 1166 & 1127 & 2 & 584
    \end{tabular}
    \caption{Overview of the simulation runs presented in this paper. Columns contain, from left to right, the name of the run; the colour used for the run in figures; the maximum FLASH grid refinement level $l_\textrm{max}$; the method for computing the FUV radiation field protoplanetary discs are exposed to; $t_*$, the time at which the first star forms; $t_{*,>7\textrm{M}_\odot}$, the time at which the first star more massive than 7 $\textrm{M}_\odot$ formed; $M_{*,>7\textrm{M}_\odot}$, the mass of the first star more massive than 7 $\textrm{M}_\odot$ to form; $t_\textrm{end}$, the time at which the run ends; the number of stars formed; the number of protoplanetary discs in the run; the number of dispersed protoplanetary discs in the run; and the number of dynamic truncations in the run. }
    \label{tab:overview}
\end{table*}

\section{Results} \label{sec:results}

All figures in this section use the colours in Table \ref{tab:overview} to indicate specific runs. For runs m1r/g to m5r/g, which share colours, we indicate which data are from the radiative and geometric runs.

\subsection{Star formation overview}

\begin{figure*}
    \centering
    \includegraphics[width=0.94\linewidth]{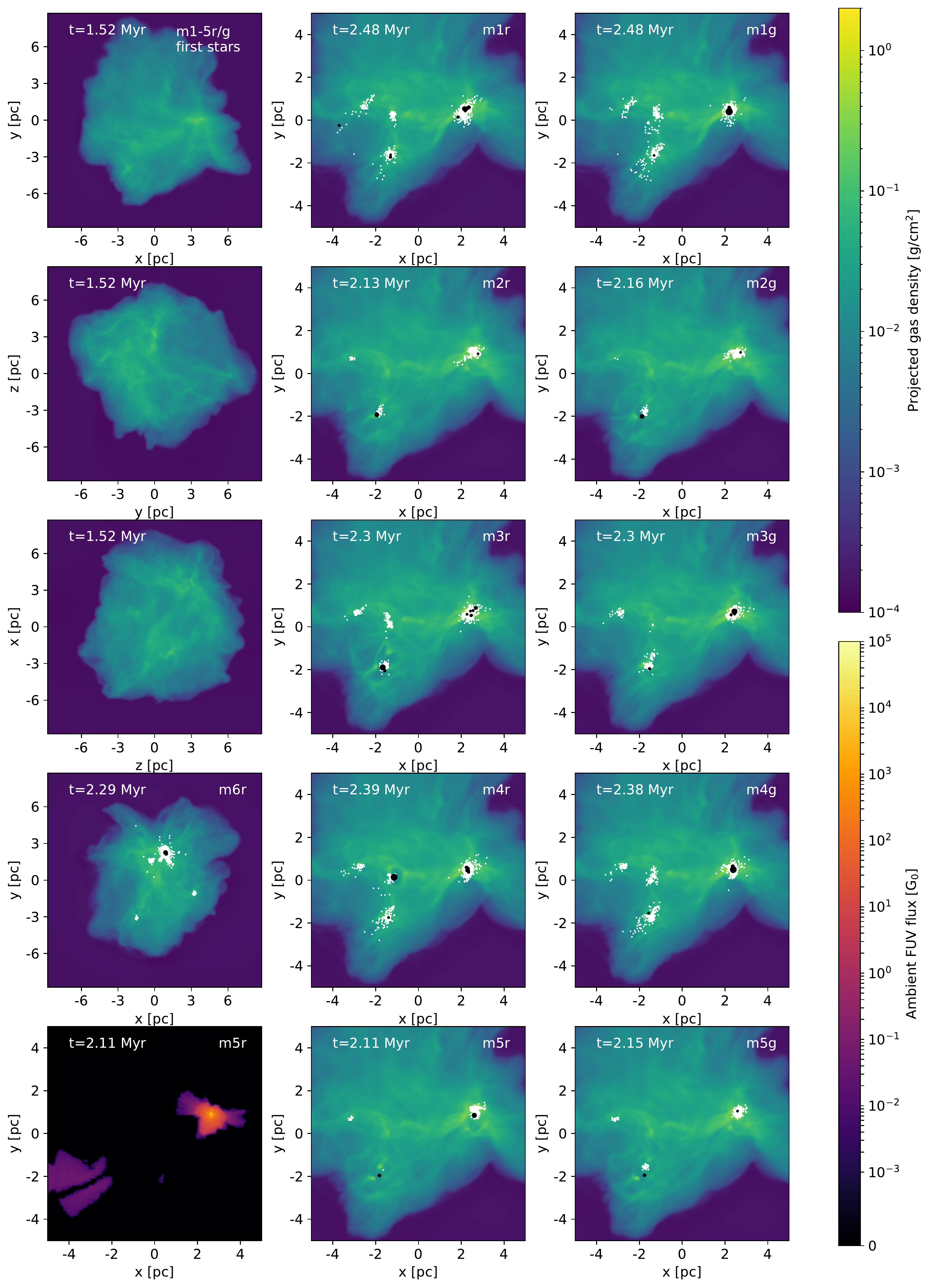}
    \caption{{\it Left column:} three projections of gas column density when star formation starts (in m1-5r/g), one at the end of m6r, and a slice of the FUV flux around the most massive star at the end of m5r. {\it Middle, right columns:} Zooms of the gas column density at the end of runs m1r-m5r and m1g-m5g, respectively. Black points are massive stars (size corresponds to mass), white points are non-massive stars.}
    \label{fig:hydro_maps}
\end{figure*}

In Fig. \ref{fig:hydro_maps}, we show a number of maps of our simulation. In the main runs, star formation has started in four dense regions, with the majority forming in the biggest overdensity on the right side of the $z$ axis projected maps. Three more subclusters form in an arc to the left (with one subcluster forming between 2.15 and 2.25 Myr). Some runs show signs of feedback bubbles being blown by massive stars, e.g. runs m5r/g in the main overdensity and in the bottom subcluster. In run m6r, there is also a subcluster containing the majority of stars and multiple smaller subclusters. 

\begin{figure}
    \centering
    \includegraphics[width=\linewidth]{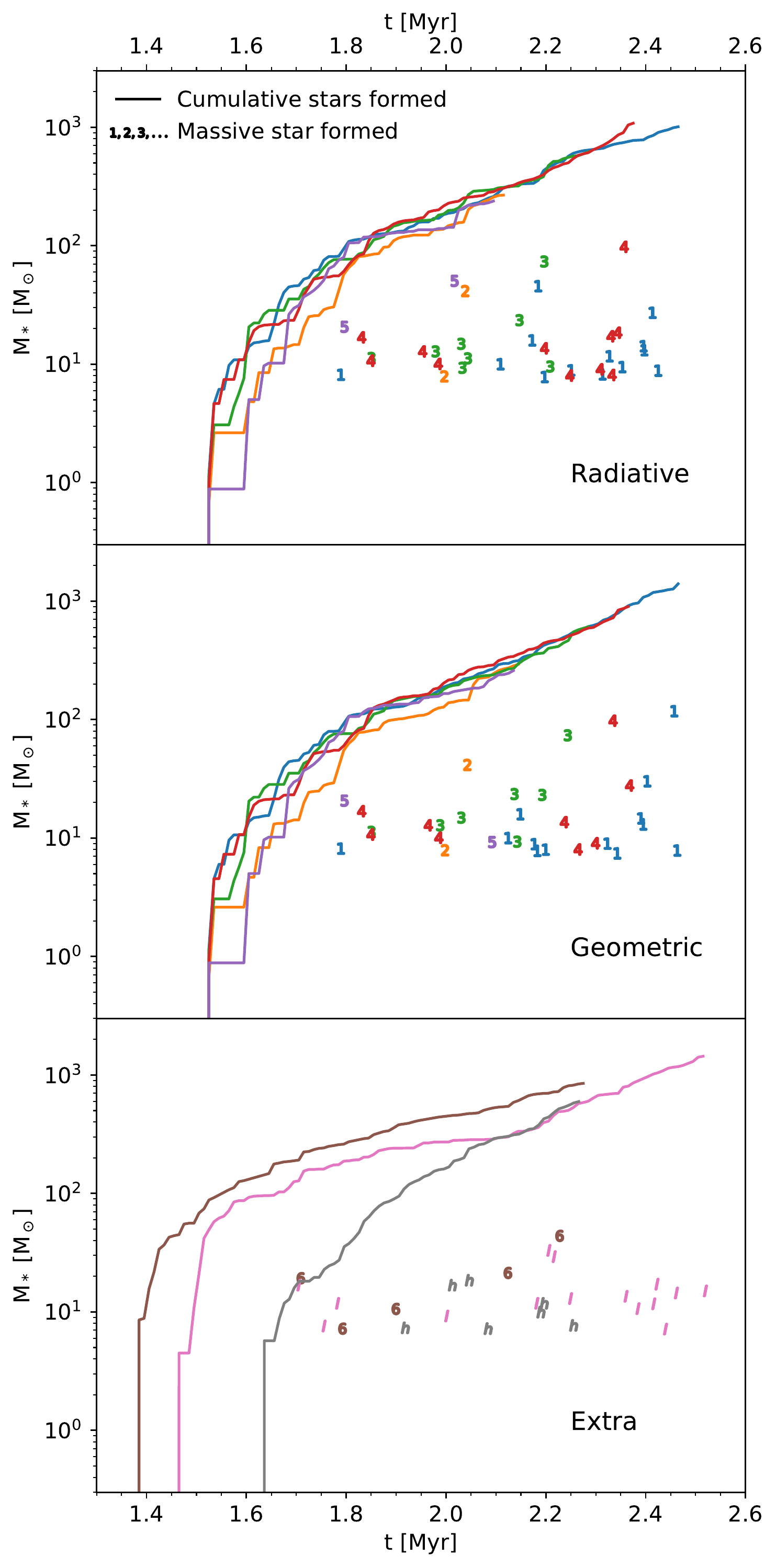}
    \caption{The star formation history of all runs. The lines indicate the cumulative mass of formed stars. Symbols (the run number for the main runs and m6r, $l$ for l3r, and $h$ for h3r) show the mass and formation time of massive stars.}
    \label{fig:sfh_overview}
\end{figure}

In Fig. \ref{fig:sfh_overview}, we show an overview of the star formation history of our simulations. As discussed in Sec. \ref{sec:overview}, the difference in EPE rate between the radiative and geometric runs has led to diverging star formation histories. For each pair of runs, the first massive star to form is the same star (with the same mass, at the same time), and some later massive stars also form at the same time (e.g. the two born in runs m4r and m4g up to 0.1 Myr after the first star formed). However, in other runs entirely different massive stars form (e.g. the two $\lesssim10 \textrm{ M}_\odot$ stars at 2 Myr in run m3r, which did not form in m3g). There are also large differences in the star formation history between runs l3r, m3r, and h3r, which use the same IMF seed and cloud initial conditions but different resolution. In general, star formation starts earlier at lower resolution. The massive stars that form are also entirely different. This makes it so that these runs are not a straight-forward test of numerical convergence. 

\begin{figure}
    \centering
    \includegraphics[width=\linewidth]{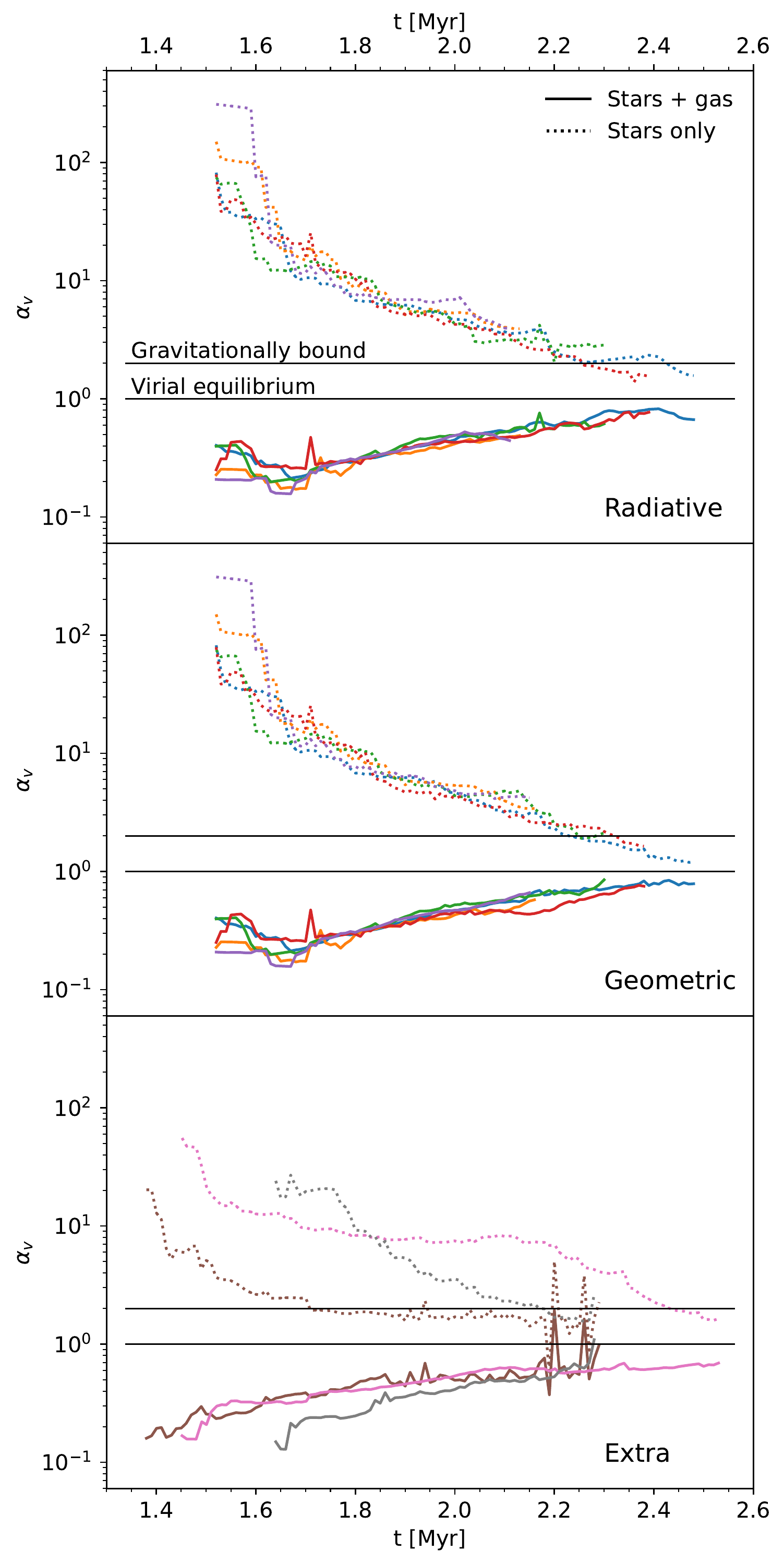}
    \caption{The evolution of the stellar virial ratio of all runs. Solid lines include the gravitational potential of the gas and stars, dotted lines only that of the stars. The horizontal black lines mark virial ratios of 1 (virial equilibrium) and 2 (gravitational boundedness).}
    \label{fig:dynamical_state}
\end{figure}

In Fig. \ref{fig:dynamical_state}, we show the time evolution of the virial ratio of the stellar population of all runs. Each cluster typically starts subvirial when the potential of both stars and gas is counted, but globally unbound if only the stellar potential is considered. Both virial ratios evolve toward a value of 1. The stars become globally bound after 1 Myr even if the residual gas were to be removed instantaneously. This trend is mostly due to the stellar potential becoming more prominent; the stellar kinetic energy and the gas potential remain of comparable magnitude.

\begin{figure}
    \centering
    \includegraphics[width=\linewidth]{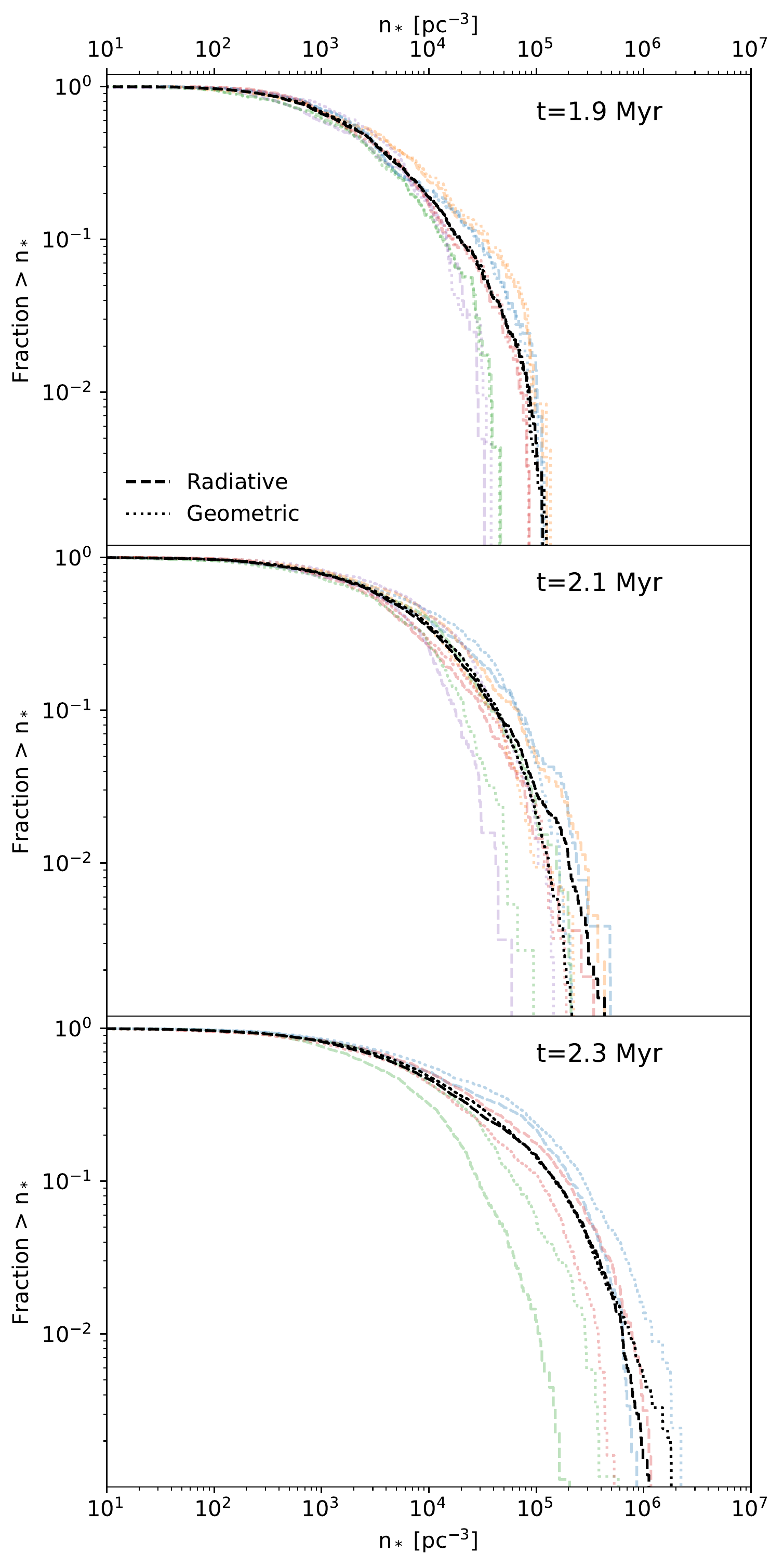}
    \caption{The fraction of stars in a local stellar density greater than some value, at three moments in time, for the main radiative and geometric runs. Black lines are aggregated over all simulations that have data at that moment.}
    \label{fig:cdf_local_density}
\end{figure}

In Fig. \ref{fig:cdf_local_density}, we show the distribution of all stars' local stellar density. Each panel shows the distribution at a different moment in time. An individual star's local stellar density is calculated using the unbiased density estimator of \citet{Casertano1985}, using the distance to the 6$^\mathrm{th}$ nearest neighbour. The distribution evolves to slightly larger values through time; the median density increases from $\sim$10$^3$ pc$^{-3}$ to $\sim$10$^4$ pc$^{-3}$, and the maximum density increases from $\sim$10$^5$ pc$^{-3}$ to $\sim$10$^6$ pc$^{-3}$. Assuming a mean stellar mass of 0.35 M$_\odot$, the typical dynamical timescale is $\lesssim$1 Myr. There is no clear trend in the stellar density distribution between radiative and geometric runs.

\subsection{Disc evolution overview}

\begin{figure}
    \centering
    \includegraphics[width=\linewidth]{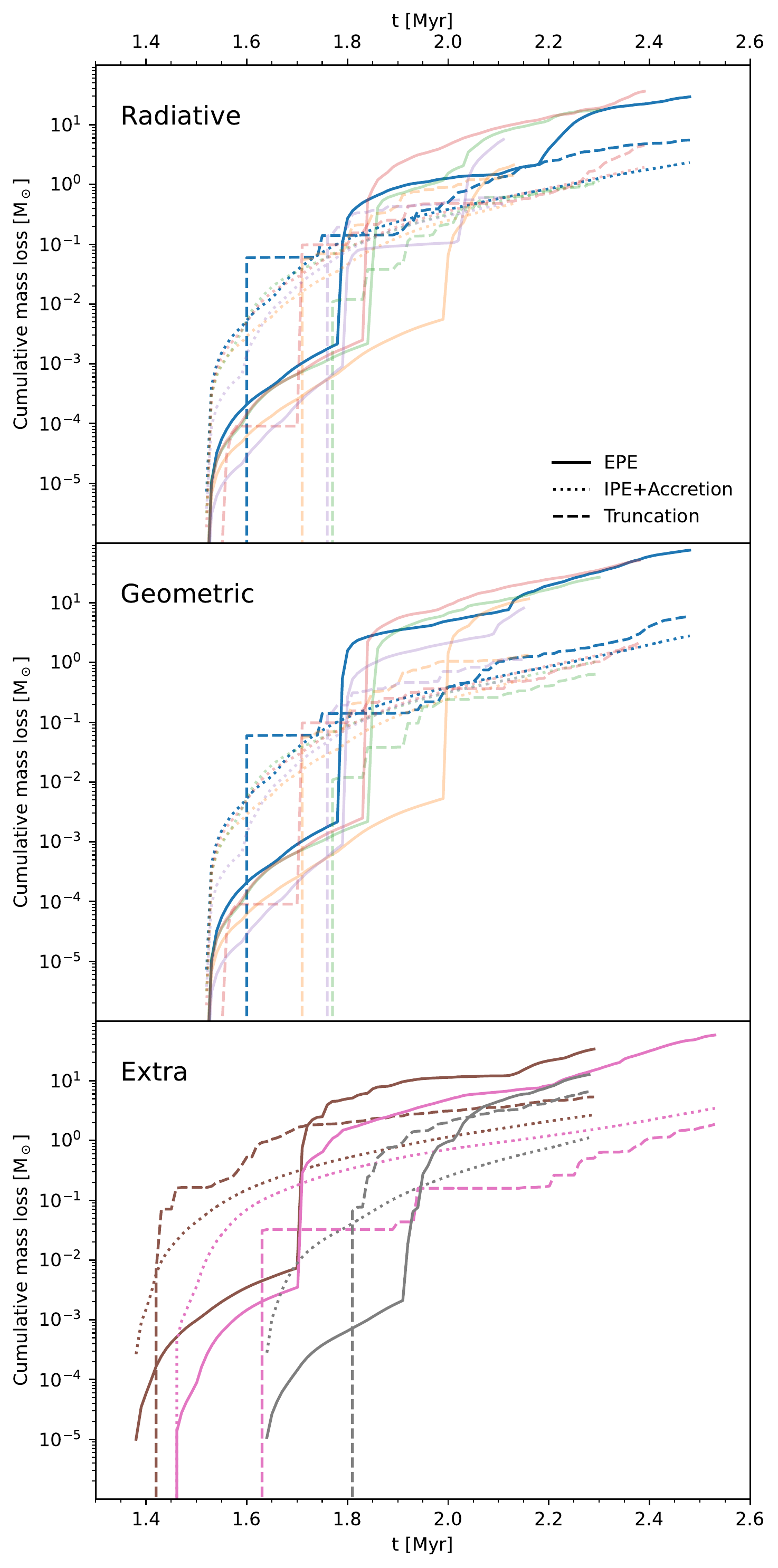}
    \caption{The cumulative mass loss through time for all runs, through different mass loss channels. These channels are external photoevaporation (EPE), internal photoevaporation (IPE) plus accretion, and dynamic truncation. In the top two panels, we emphasise runs m1r and m1g as representative examples for clearer viewing.}
    \label{fig:cumul_massloss}
\end{figure}

In Fig. \ref{fig:cumul_massloss}, we show the evolution of the total mass loss from discs through different mass loss processes, integrated over all discs for each run. The top panel shows this for the main radiative runs, the middle panel for the main geometric runs, and the bottom panel for the extra runs. Initially, most mass is lost through internal processes (i.e., IPE and accretion). At this moment, no massive stars have formed and each disc only experiences a negligible non-zero EPE mass loss rate. The first dynamic truncation events are visible as instantaneous increases in mass loss through truncation.

In each run, at the same moment the first massive star forms, the mass loss through EPE sharply increases. The mass loss then settles into a more shallow increase. This is due to the shrinking of discs; when the EPE rate is low, discs are able to expand viscously, and the outer edges of these large discs are weakly bound. When a massive star forms and the radiation field increases, this far-out mass is quickly stripped away. As the disc radius decreases, the mass loss rate also decreases. 

At late times, EPE is the dominant mass loss process in all runs. In the radiative runs, mass loss through truncation is either comparable to or larger than the combined internal processes (except in run l3r). In the geometric runs mass loss through truncation can also be lower than through internal processes.

\subsection{Effects of extinction due to cluster gas}

\begin{figure}
    \centering
    \includegraphics[width=1\linewidth]{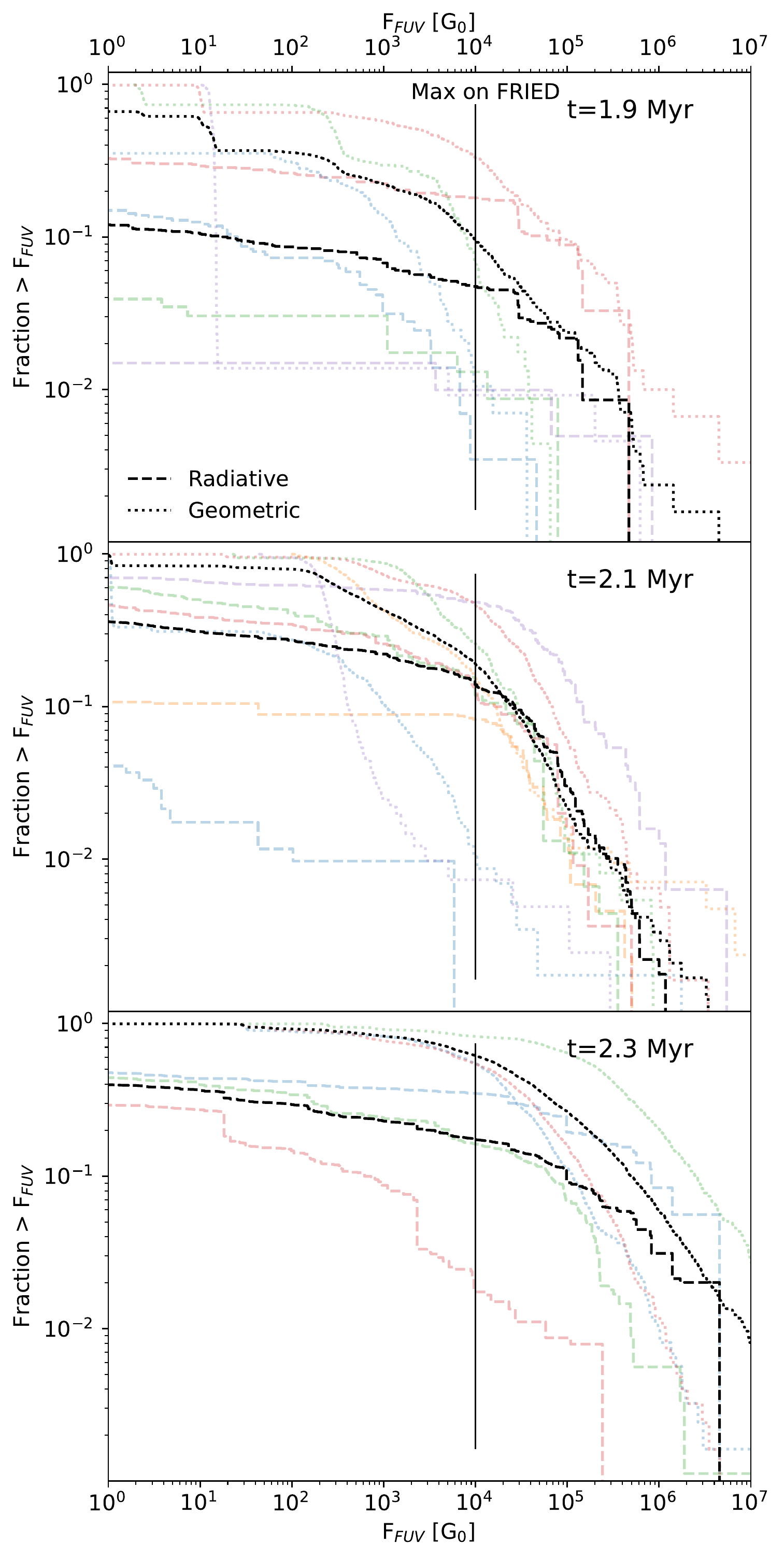}
    \caption{The fraction of discs exposed to a radiation field greater than some value, at three moments in time, for the main radiative and geometric runs. Black lines are aggregated over all simulations that have data at that moment.}
    \label{fig:cdf_radfield}
\end{figure}

In Fig. \ref{fig:cdf_radfield}, we show the distribution of FUV radiation fields to which the population of discs is exposed. Each panel shows the distribution at a different moment in time. The range of radiation fields extends beyond $10^4$ G$_0$, which is the largest value on the FRIED grid. If a disc is exposed to a radiation field larger than $10^4$ G$_0$, interpolation is not possible, and we instead conservatively use the EPE rate of the closest (in logarithmic space) grid point. This makes $10^4$ G$_0$ the effective maximum radiation field. 

The tail of the distribution at strong radiation ($\gtrsim$10$^{4}$ G$_0$) fields is similar for radiative and geometric runs. At 1.9 Myr, when all runs have the same population of massive stars, this holds for individual paired runs. At 2.1 Myr, when the paired runs have different populations of massive stars, the distributions integrated over all runs are still similar. At 2.3 Myr, the distributions have diverged more, but so have the star formation history. High radiation fields are found near massive stars, where the extinction can be low due to short path lengths or clearing of nearby cluster gas. The absence of radiation fields $>10^4$ G$_0$ on the FRIED grid thus has comparable impact on radiative and geometric runs.

At lower radiation fields, the distributions differ. In radiative runs, more discs experience radiation fields in the range $1-10^4$ G$_0$ than in geometric runs. We conclude that extinction due to cluster gas mostly shields discs in moderate to high radiation fields, while there remains a minority of discs exposed to very high radiation fields. Still, as late as $\sim$0.5 Myr after the formation of the first massive star, about half the discs in radiative runs are exposed to radiation fields below the mean interstellar level. 

\begin{figure}
    \centering
    \includegraphics[width=1\linewidth]{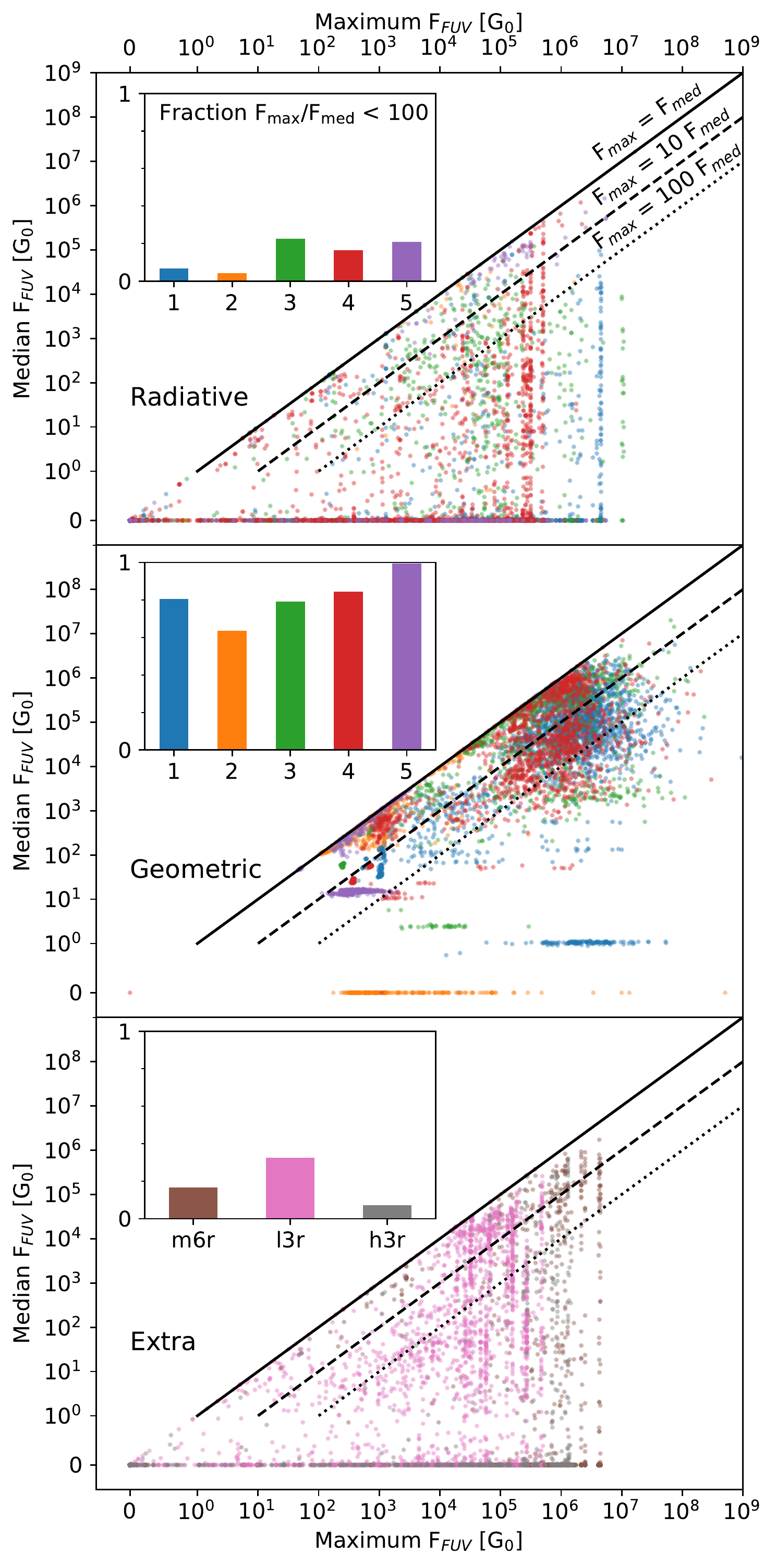}
    \caption{The median and maximum FUV radiation field of all discs in our main radiative (top), geometric (middle), and extra (bottom) runs. Black lines indicate a maximum radiation field equal to the median (solid), 10 times the median (dashed), and 100 times the median (dotted). Also shown is the fraction of discs with a maximum-to-median ratio smaller than 100. Note that the axes are logarithmic above 1 G$_0$ and linear below that.}
    \label{fig:radfield_max_vs_median}
\end{figure}

In Fig. \ref{fig:radfield_max_vs_median}, we show for all runs the distribution of the median and maximum radiation fields to which discs have been exposed. This demonstrates the variability of the radiation field. On the diagonal, a disc has been exposed to a constant radiation field. 

In general, discs in radiative runs have larger variations in radiation field when compared to geometric runs. While the maximum flux is similar in radiative and geometric runs, the median fluxes are much lower in the radiative runs. In radiative runs, the majority of discs have ratios of maximum to median flux $>100$. In geometric runs, the majority has a ratio $<100$. The trend can be understood by cluster gas creating very large spatial gradients in the radiation field. In the radiative runs, a disc can move across a ridge of gas, quickly going from low to high radiation field or the other way around. In the geometric runs, the radiation field varies at most with the inverse square law. 

\begin{figure}
    \centering
    \includegraphics[width=1\linewidth]{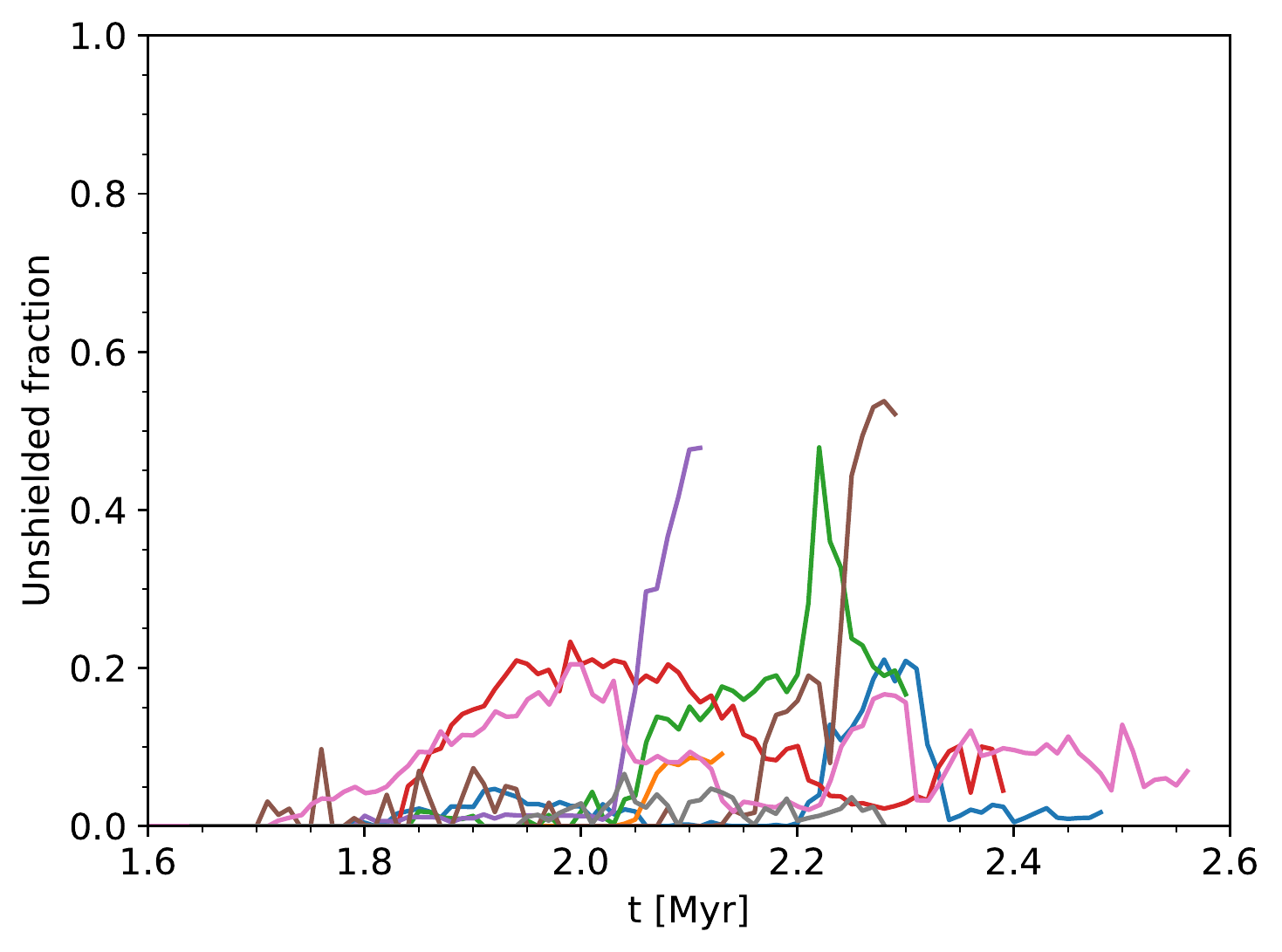}
    \caption{The fraction of discs in our radiative runs that are unshielded in the FUV, meaning that the radiation they received has been extincted by a net optical depth of unity or less.}
    \label{fig:unshielded_fraction}
\end{figure}

In Fig. \ref{fig:unshielded_fraction}, we show the fraction of discs in our radiative runs (including l3r, h3r, and m6r) that are effectively unshielded. We define this by first computing the ratio of the (extincted) radiation field to the geometric radiation field, and selecting those with a ratio $>e^{-1}$ (i.e. an optical depth of at most unity if there were only a single radiation source). This fraction represents the degree of global gas clearing. The unshielded fraction is not a monotonically increasing quantity on the timescale of our runs, and does not becomes greater than 60\% as late as $\sim$0.5 Myr after the formation of the first massive star. The results for run m6r do not differ notably from the other runs.

\begin{figure}
    \centering
    \includegraphics[width=1\linewidth]{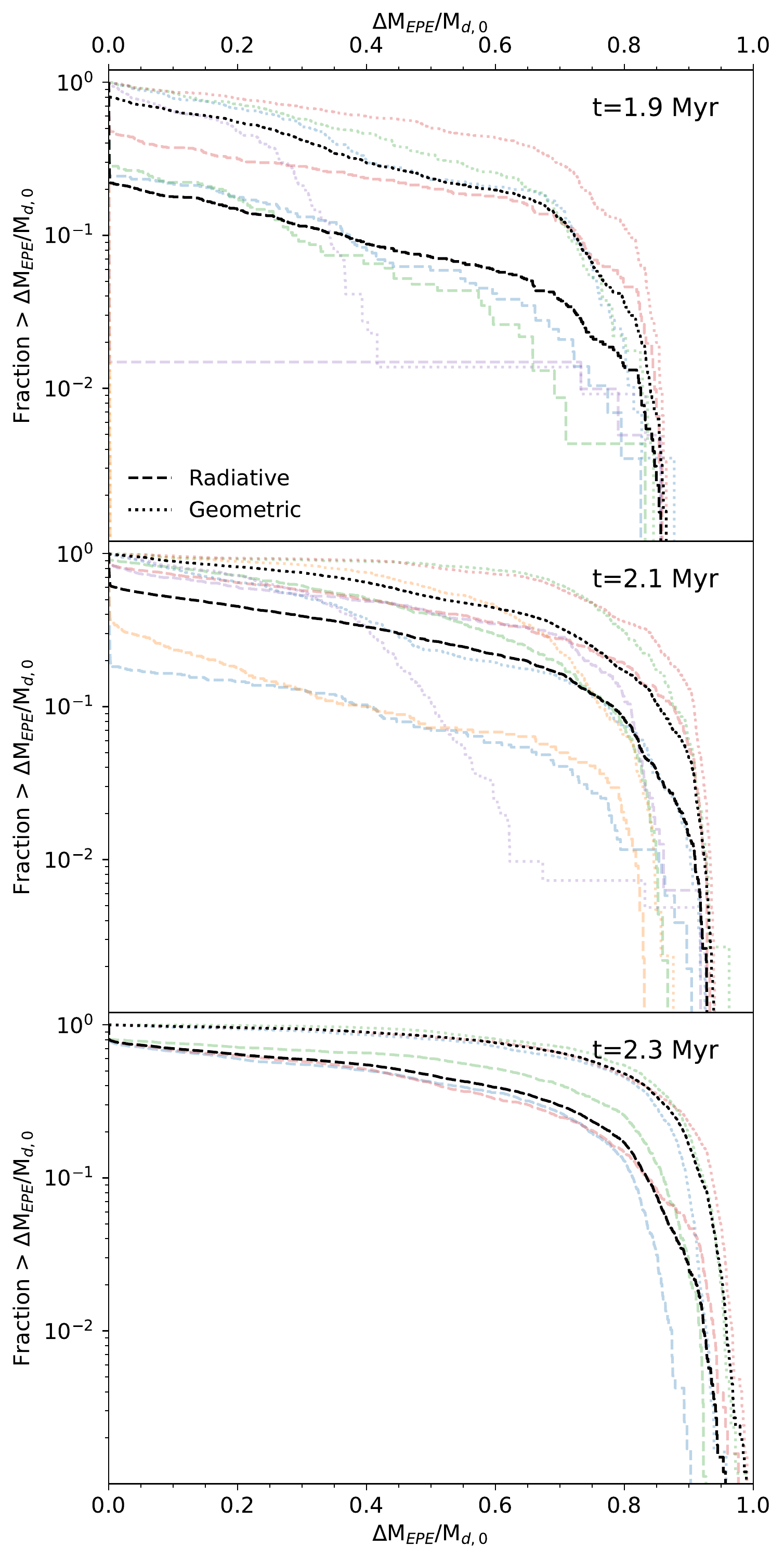}
    \caption{The fraction of discs that have lost mass (normalised to their initial disc mass) due to EPE greater than some value, at three moments in time, for the main radiative and geometric runs. Black lines are aggregated over all simulations that have data at that moment.}
    \label{fig:cdf_epe_massloss}
\end{figure}

In Fig. \ref{fig:cdf_epe_massloss}, we show the distribution of disc mass loss through EPE, normalised to the initial disc mass. Each panel shows the distribution at a different moment in time.

Between runs with the same IMF seed and different radiation field methods, discs in the geometric runs have lost more mass through EPE than in the radiative runs. The exception to this is runs m5r and m5g, both at 1.9 and 2.1 Myr. At 1.9 Myr, the radiative and geometric distributions are similar at mass losses $>0.4$, but the radiative distribution is slightly above the geometric one. This implies that the same stars are subject to EPE, but the total number of stars differs. The first massive star in these runs forms in a small subcluster, and the large subcluster is initially fully shielded. At 2.1 Myr, the radiative distribution is above the geometric one for most mass losses. This is due to the formation of a $\sim$50 M$_\odot$ star at 2 Myr in the radiative run that is absent in the geometric run. Compared to other radiative runs, m5r is not an outlier at 2.1 Myr.

At 2.3 Myr, $\sim$0.5 Myr after the formation of the first massive star, almost all discs in the geometric runs have lost more than half their mass through EPE, compared to about half in the radiative runs. The tails at high mass losses also differ considerably. 10-30\% of discs in geometric runs lose $>$90\% of their mass through EPE, compared to about 3\% of discs in radiative runs. Contrast this with the similarity of the tail at high radiation fields. This can be explained through the radiation field variability demonstrated in Fig. \ref{fig:radfield_max_vs_median}. Discs that occupy the tail at high radiation fields do not do so all of the time, both in radiative and geometric runs. The lower median radiation field in radiative runs appears to lead to a lower mass loss integrated over time.

\subsection{Dynamic truncations}

\begin{figure}
    \centering
    \includegraphics[width=1\linewidth]{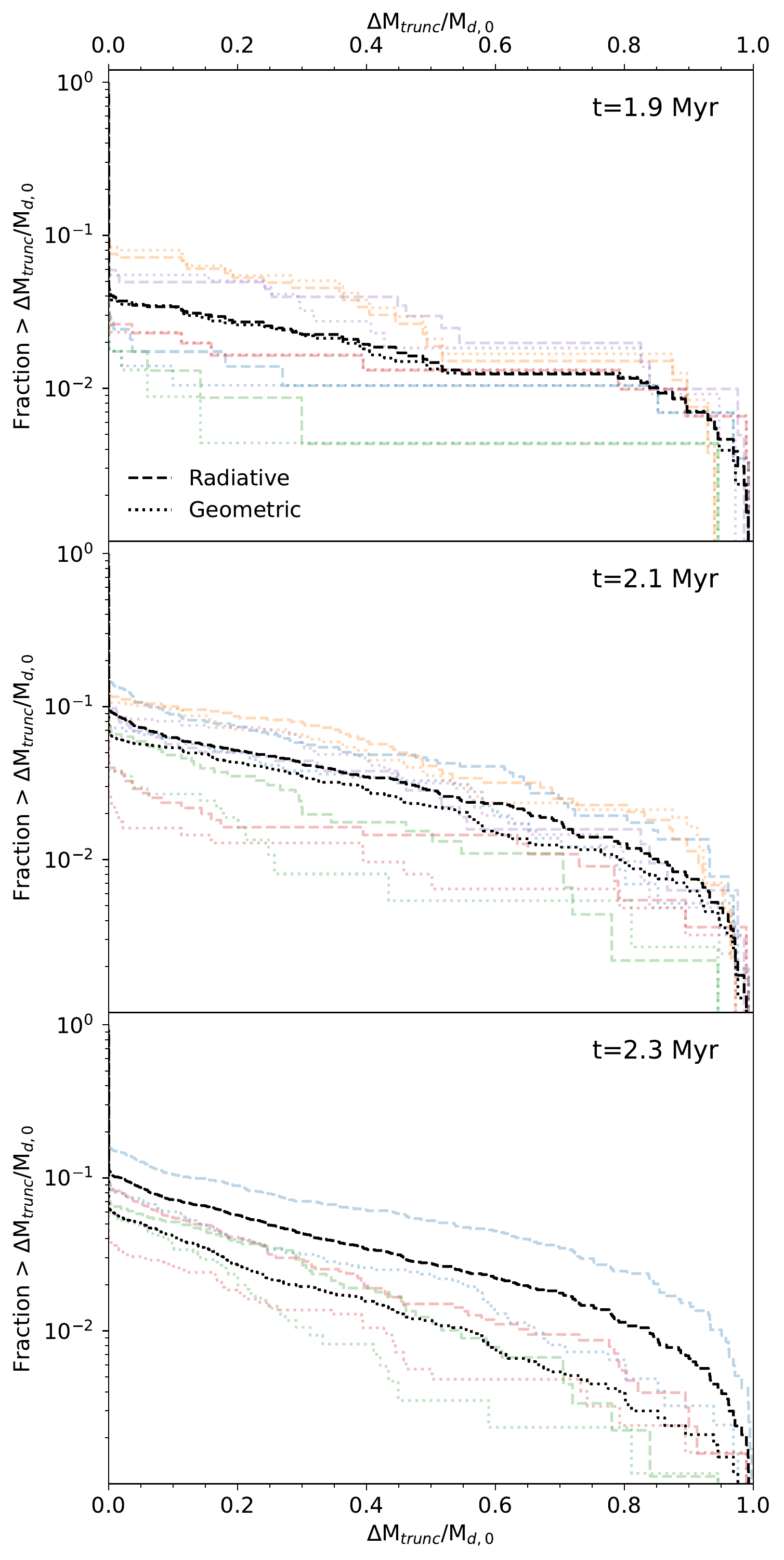}
    \caption{The fraction of discs that have lost mass (normalised to their initial disc mass) due to dynamic truncation greater than some value, at three moments in time, for the main radiative and geometric runs. Black lines are aggregated over all simulations that have data at that moment.}
    \label{fig:cdf_trnc_massloss}
\end{figure}

In Fig. \ref{fig:cdf_trnc_massloss}, we show the distribution of disc mass loss through dynamic truncations, normalised to the initial disc mass. Each panel shows the distribution at a different moment in time.

Until the formation of the first massive star (due to the different treatments of EPE) the radiative and geometric runs evolve identically (up to numerical round-off errors). At 1.9 Myr, the runs start to diverge, and the stellar orbits are still almost identical, resulting in identical truncation events. This leads to similar truncation mass loss distributions.

As the models evolve further and the paired runs diverge, different truncations happen, and the mass loss distributions diverge. In general, discs in radiative runs lose more mass through truncation than discs in geometric runs, especially when comparing paired runs. This is a direct consequence of EPE. Both truncation and EPE act on the outer disc. A disc that has lost material through EPE will be smaller, and can lose less mass through truncation. In an extreme case a certain encounter may not truncate a small, photoevaporated disc, while it may have truncated a larger disc. The opposite case can also happen, where a dynamically truncated disc has a lower EPE rate due to the remaining material being deeper in the host star's potential well. This further reduces EPE in radiative runs, in addition to lower radiation fields, but not to a point where dynamic truncation globally dominates over EPE in terms of mass loss.

As also seen in Fig. \ref{fig:cumul_massloss}, EPE mass loss dominates truncation mass loss, both in the radiative and geometric runs. However, a small fraction of discs lose a large amount of mass though truncation. At 2.3 Myr, about 0.8 Myr after the first star forms, about 3\% of discs in radiative runs and 1\% of discs in geometric runs lose more than half their initial mass through dynamic truncations. The fraction of discs that lose a non-negligible amount of mass through truncations is about 10\% in radiative runs and 6\% in geometric runs. A  very small fraction of discs, $<$1\%, loses $>$90\% of their mass through truncations, even at early times.

\begin{figure}
    \centering
    \includegraphics[width=1\linewidth]{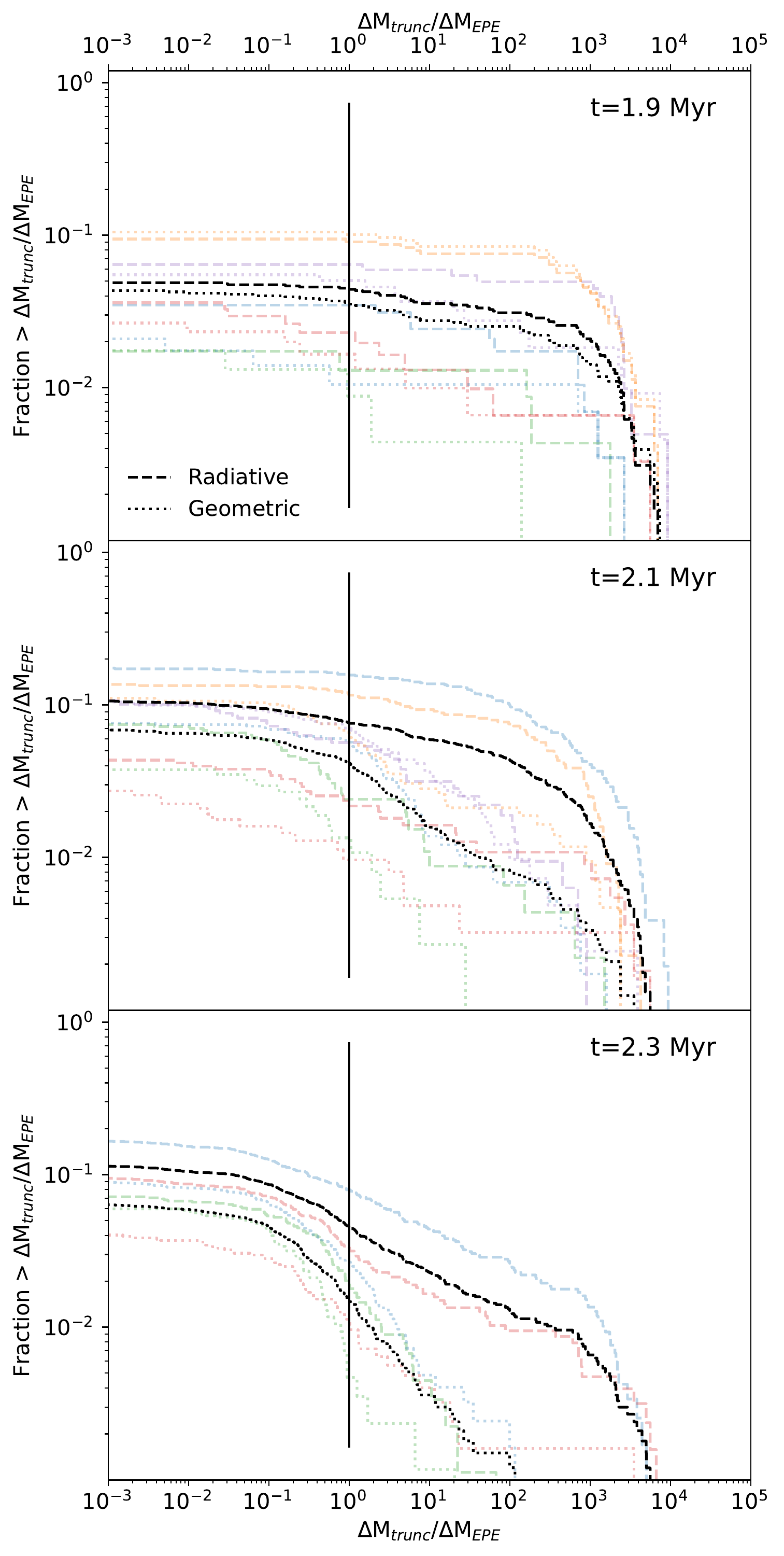}
    \caption{The fraction of discs for which the ratio of mass loss due to truncation to mass loss through EPE is greater than some value, at three moments in time, for the main radiative and geometric runs. Black lines are aggregated over all simulations that have data at that moment. The black vertical line indicates a ratio of one.}
    \label{fig:cdf_ratio_massloss}
\end{figure}

In Fig. \ref{fig:cdf_ratio_massloss}, we show the distribution of the ratio of disc mass loss through truncations to that lost through EPE. Each panel shows the distribution at a different moment in time. 

While most discs lose more mass through EPE than through truncations, a considerable fraction of discs ($\sim$1-10\%) are dominated by truncation mass loss. This fraction is greater for the radiative runs. This shows that EPE may be the dominant mass loss process globally, but truncation can be the dominant mass loss process for a minority of discs.

\begin{figure}
    \centering
    \includegraphics[width=1\linewidth]{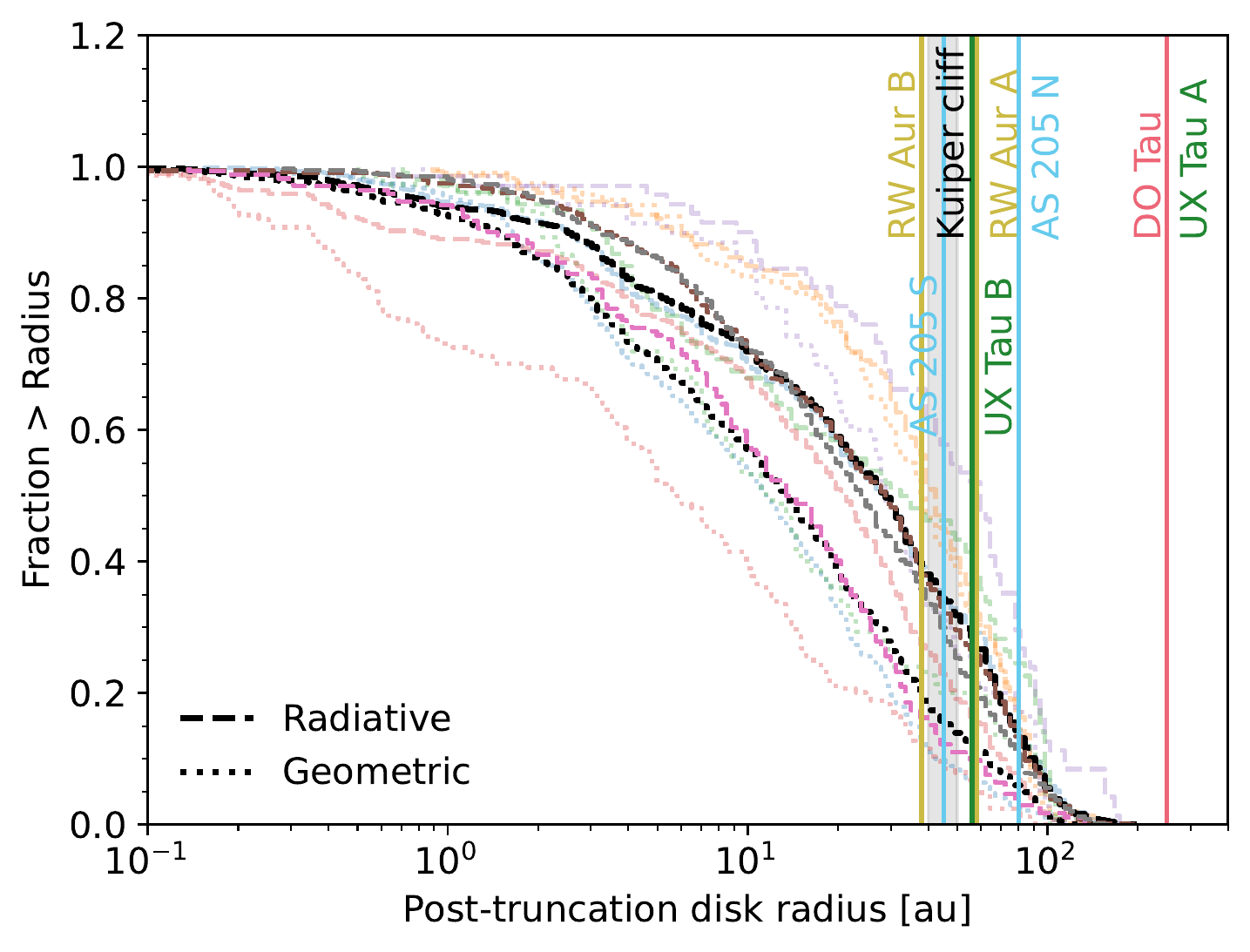}
    \caption{The fraction of truncation events that truncate a disc to a radius greater than some value. Black lines are aggregated over simulations m1r/g-m5r/g. The grey shaded region indicates the approximate location of the Kuiper cliff \citep{Allen2001,PortegiesZwart2018}. Coloured lines indicate the CO emission radii of AS 205 N and S \citep{Salyk2014}, RW Aurigae A and B \citep{Rodriguez2018}, UX Tau A and C \citep{Zapata2020}, and DO Tau \citep{Huang2022}.}
    \label{fig:trnc_radii}
\end{figure}

In Fig. \ref{fig:trnc_radii}, we show the distribution of disc radii after truncation in all simulations. This distribution is related to the distribution of close encounters between stars, but adjusted by stellar masses (Eq. \ref{eq:trnc}) and the disc radius distribution. A given close encounter only results in a truncation if a disc is larger than the truncation radius, so the truncation radius distribution of a population of smaller discs will be skewed to smaller radii. 

We also show the approximate region of the Kuiper cliff at 40-50 au \citep{Allen2001, PortegiesZwart2018}. The Kuiper cliff is within the distribution of truncation radii, but well above the median, which is at 10-30 au. \citet{Allen2001} already suggested that a truncation or excitation event might have been the origin of the Kuiper cliff, and our simulations imply that such truncation events can reasonably happen in high-density star forming regions. 

We also show the radii (measured in CO emission) of a number of discs observed to have been in dynamic encounters: AS 205 N and S \citep[80 and 45 au,][]{Salyk2014}; RW Aurigae A and B \citep[58 and 38 au,][]{Rodriguez2018}; UX Tau A and C \citep[250 and $<$56 au,][]{Zapata2020}; and DO Tau \citep[250 au,][]{Huang2022}. We disregard the discs of HV Tau and Z CMA AB because they are circumbinary discs, which we do not model. For Z CMA C, no CO radius was given.

All observed post-truncation radii fall above the median post-truncation radii in most of our simulations. Except for DO Tau and UX Tau A, all fall within the $90^\textrm{th}$ percentile of the majority of simulations. These two exceptions are likely due to discs with radii $>$250 au not being present in our simulations due to strong EPE. We note that the observed discs are all in low mass star forming regions lacking OB stars.

\subsection{Disc lifetime} \label{sec:lifetime}

Our simulations cover such a short period of time that there are few fully dispersed discs (at most $\sim$1\%, Table \ref{tab:overview}, second to last column; a disc is considered dispersed when its gas mass is less than $8\cdot 10^{-5}$ M$_\odot$), so we are unable to estimate the disc lifetime via the decrease of the fraction of stars with discs. In addition, the still ongoing formation of stars with discs complicates such an analysis. Therefore we investigate the time it has taken discs to lose a certain fraction of mass.

\begin{figure*}
    \centering
    \includegraphics[width=1\linewidth]{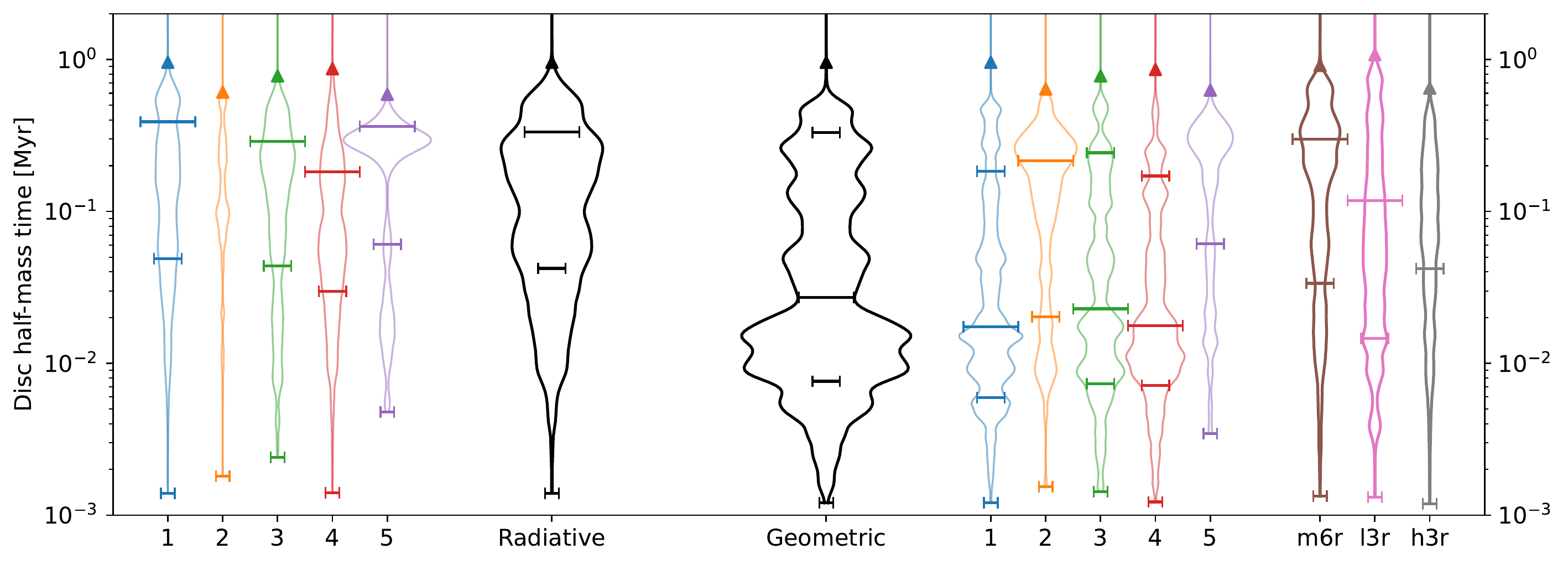}
    \caption{Violin plots of the half-mass time distributions of all simulation runs, plus those aggregated over all main radiative and geometric runs. Horizontal dashes mark the 0$^\textrm{th}$, 16$^\textrm{th}$, 50$^\textrm{th}$, 84$^\textrm{th}$, and 100$^\textrm{th}$ percentiles (width marks different percentiles; the 0$^\textrm{th}$ and 100$^\textrm{th}$ are narrowest, the 50$^\textrm{th}$ is widest). Upward triangles denote the maximum disc age in a run, which forms an upper limit on the half-mass time during the simulation time. For this plot we consider the half-mass times of discs that retain more than half their mass as infinite (which means that some percentiles can be at infinity).}
    \label{fig:disc_halflife}
\end{figure*}

In Fig. \ref{fig:disc_halflife} we show the distributions of the half-mass times of the discs in all simulation runs. If a disc has lost less than half its mass by the end of the simulation, we take its half-mass time to be infinity. Horizontal dashes indicate (from bottom to top) the 0$^\textrm{th}$, 16$^\textrm{th}$, 50$^\textrm{th}$, 84$^\textrm{th}$, and 100$^\textrm{th}$ percentiles. The upward triangles indicate the maximum age of discs in each simulation, which is the upper limit of half-mass times during the simulation time. We consider the half-mass time of discs that retain more than half their mass to be infinite, although after further evolution they could be any value.

For most sets of paired runs, the half-mass times of the geometric runs are shorter than those of the radiative runs. The exception is runs m5r and m5g, which can be attributed to the formation of the $\sim$50 M$_\odot$ in m5r. Aggregated over all main radiative and geometric runs, the median half-mass time of the geometric runs is smaller than the radiative runs' 16$^\textrm{th}$ percentile. Even though dynamic truncation is more efficient in the radiative runs than the geometric runs, the net effect of shielding in those runs is slower mass loss per disc. The half-mass time distribution of run m6r (which had a different realisation of the cloud's initial velocity field) is similar to most of the main radiative runs except m2r, implying that the realisation of the IMF is more important than the realisation of the (velocity field of the) cloud. 

The half-mass times are difficult to compare to observed disc lifetimes. IPE and accretion rates decrease with time in our model, and EPE rates decrease with disc mass and radius, although the radiation field can increase as more stars form. In general, the disc lifetime is not simply twice the half-mass time. Still, it is striking how short these timescales are; in all but 3 out of 13 runs, the median half-mass time is smaller than 0.5 Myr. In 3 runs, it is even $<0.05$ Myr. These values are about an order of magnitude or more smaller than disc lifetimes estimated from disc fractions, $\sim$2.5-8 Myr \citep[e.g.][]{Haisch2001,Richert2018,Michel2021}. However, it should be noted that those values are derived from disc populations in young stellar associations with a wide variety of masses, including those that lack massive stars. In some of our runs, radiation shielding increased disc half-mass times by an order of magnitude, so an increase of similar magnitude in regions lacking massive stars appears plausible. On the other hand, \citet{Pfalzner2022} recently found that disc lifetimes in dense clusters can be as low as 0.77 Myr, which is of similar magnitude to half-mass times in our radiative runs.

\begin{figure}
    \centering
    \includegraphics[width=1\linewidth]{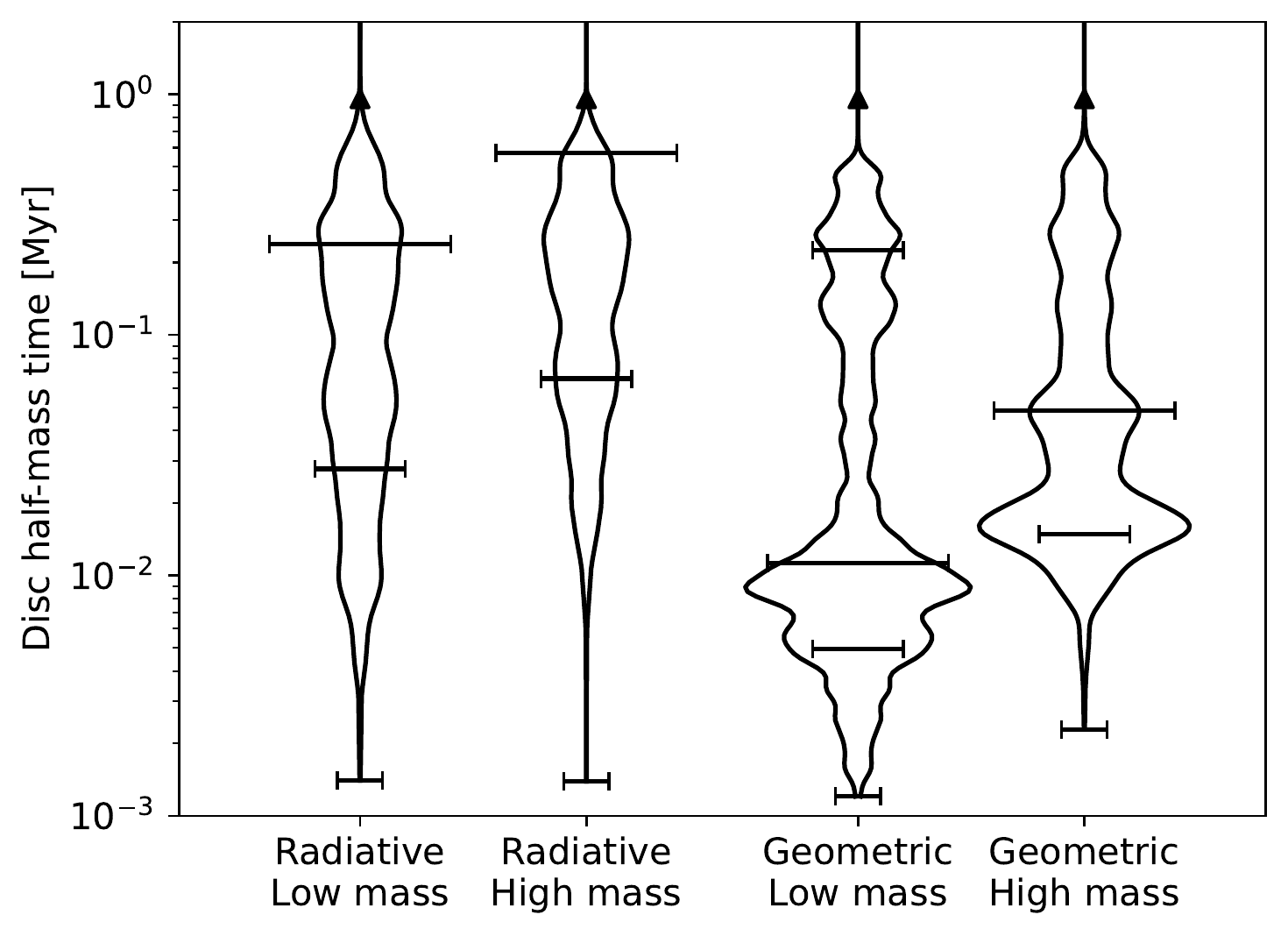}
    \caption{Violin plots of the half-mass time distribution of the aggregated main radiative and geometric runs, split into the lower and upper 50$^\textrm{th}$ percentiles of host star mass (the median is $\sim$0.25 M$_\odot$). Horizontal dashes mark the 0$^\textrm{th}$, 16$^\textrm{th}$, 50$^\textrm{th}$, 84$^\textrm{th}$, and 100$^\textrm{th}$ percentiles (width marks different percentiles; the 0$^\textrm{th}$ and 100$^\textrm{th}$ are narrowest, the 50$^\textrm{th}$ is widest). Upward triangles denote the maximum disc age in a run, which forms an upper limit on the half-mass time during the simulation time. For this plot we consider the half-mass times of discs that retain more than half their mass as infinite (which means that some percentiles can be at infinity)}
    \label{fig:hm_time_mass_split}
\end{figure}

In Fig. \ref{fig:hm_time_mass_split} we show the half-mass times of the lower and upper 50$^\textrm{th}$ percentiles of host star mass (the median is $\sim$0.25 M$_\odot$), aggregated over all main radiative and geometric runs. For both radiative and geometric runs, the median half-mass time of the low mass population is smaller than that of the high mass population. This is in apparent conflict with the results of e.g. \citet{Carpenter2006,Dahm2007,Ribas2015}, who find lower disc fractions for higher mass stars in star forming regions including those with massive stars. However, these results are obtained for older systems, and we don't resolve the high end of the lifetime distribution. At later times mass segregation may migrate higher mass stars closer to the region's massive stars than lower mass stars. Also, the ratio between median half-mass time of the high and low mass populations is smaller for radiative runs ($\sim$2) than for geometric runs ($\sim$4). Additionally, in the geometric runs the low mass median half-mass time is smaller than the high mass' 16$^\textrm{th}$ percentile, while in the radiative runs this low mass median is larger than the high mass 16$^\textrm{th}$ percentile. These results are consistent with the picture of \citet{Wilhelm2022}, who found that EPE has a larger impact on the lifetime of discs around low mass stars than on that of discs around high mass stars. This difference is caused by a combination of EPE being more effective in the loosely-bound discs around lower mass stars, and of accretion and IPE being more effective in discs around higher mass stars. Overall, radiation shielding appears to slightly relax the tension with the observations, even if it doesn't resolve it.

\subsection{Planet-forming potential}

The dust model of \citet{Haworth2018a} describes how dust is removed from a disc by EPE, but is agnostic to the ultimate fate of the remaining dust. It effectively provides an upper limit on the solid mass available for planet formation. 

In our analysis we restrict ourselves to discs that do not lose any more solid mass (meaning that the upper limit on the amount of material available for planet formation has been established), which happens when dust grains grow to sizes too large to be entrained in the photoevaporative wind. In \citet{Haworth2018a}, this growth is implemented as an exponential decrease in the solid mass loss rate with time. The timescale of this decrease is twice the timescale of dust growth, which is

\begin{equation}
\tau_d = \frac{1}{\delta}\sqrt{\frac{R^3}{GM_*}},
\end{equation}

\noindent
where $\delta$ is the dust to gas ratio (assumed to be 0.01), $R$ is the outer disc radius, and $M_*$ is the mass of the host star. For every disc, we compute this timescale given the final disc radius. If a disc has lost at most 1\% of its solid mass over the last ten dust growth timescales, and is older than ten such timescales, we consider dust EPE to be finished. In our analysis we focus on two populations of host stars: low mass stars, which are the most common in the present-day universe, and Sun-like stars. We define these populations as stars in the range of 0.08-0.3 M$_\odot$ and 0.7-1.3 M$_\odot$, respectively. 

\begin{figure*}
    \centering
    \includegraphics[width=1\linewidth]{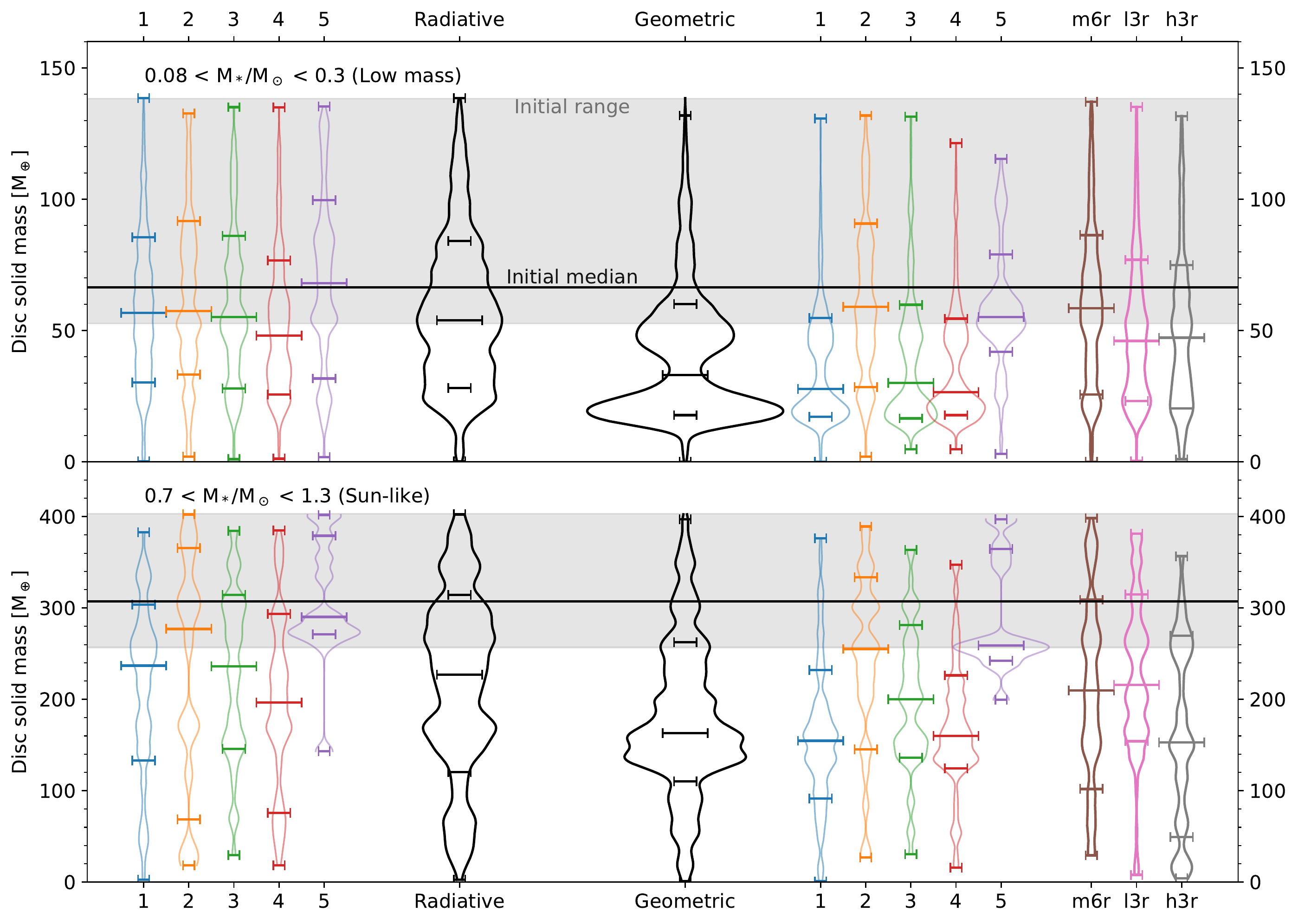}
    \caption{Violin plots of the solid mass distribution (of discs where dust EPE has finished) of all simulation runs, plus those aggregated over all main radiative and geometric runs. Host stars are in the ranges 0.08-0.3 M$_\odot$ (top panel) and 0.7-1.3 M$_\odot$ (bottom panel). Horizontal dashes mark the 0$^\textrm{th}$, 16$^\textrm{th}$, 50$^\textrm{th}$, 84$^\textrm{th}$, and 100$^\textrm{th}$ percentiles (width marks different percentiles; the 0$^\textrm{th}$ and 100$^\textrm{th}$ are narrowest, the 50$^\textrm{th}$ is widest). The grey shaded regions indicate the range of initial disc solid masses for the respective host star mass bin, the black horizontal line the median (for masses following the Kroupa IMF).}
    \label{fig:final_solid_mass}
\end{figure*}

In Fig. \ref{fig:final_solid_mass} we show the distribution of final solid masses in discs where dust EPE has finished, for all runs. Between most paired runs, discs in the radiative runs have more solid mass than those in geometric runs. The exception to this is low mass stars in runs m2r/g, although the difference is small. The median solid mass decreases from the initial value a small amount in some runs (run m5r), to a factor $\sim$2 (e.g. runs m1g, m3g, m4g). 

\begin{figure*}
    \centering
    \includegraphics[width=0.88\linewidth]{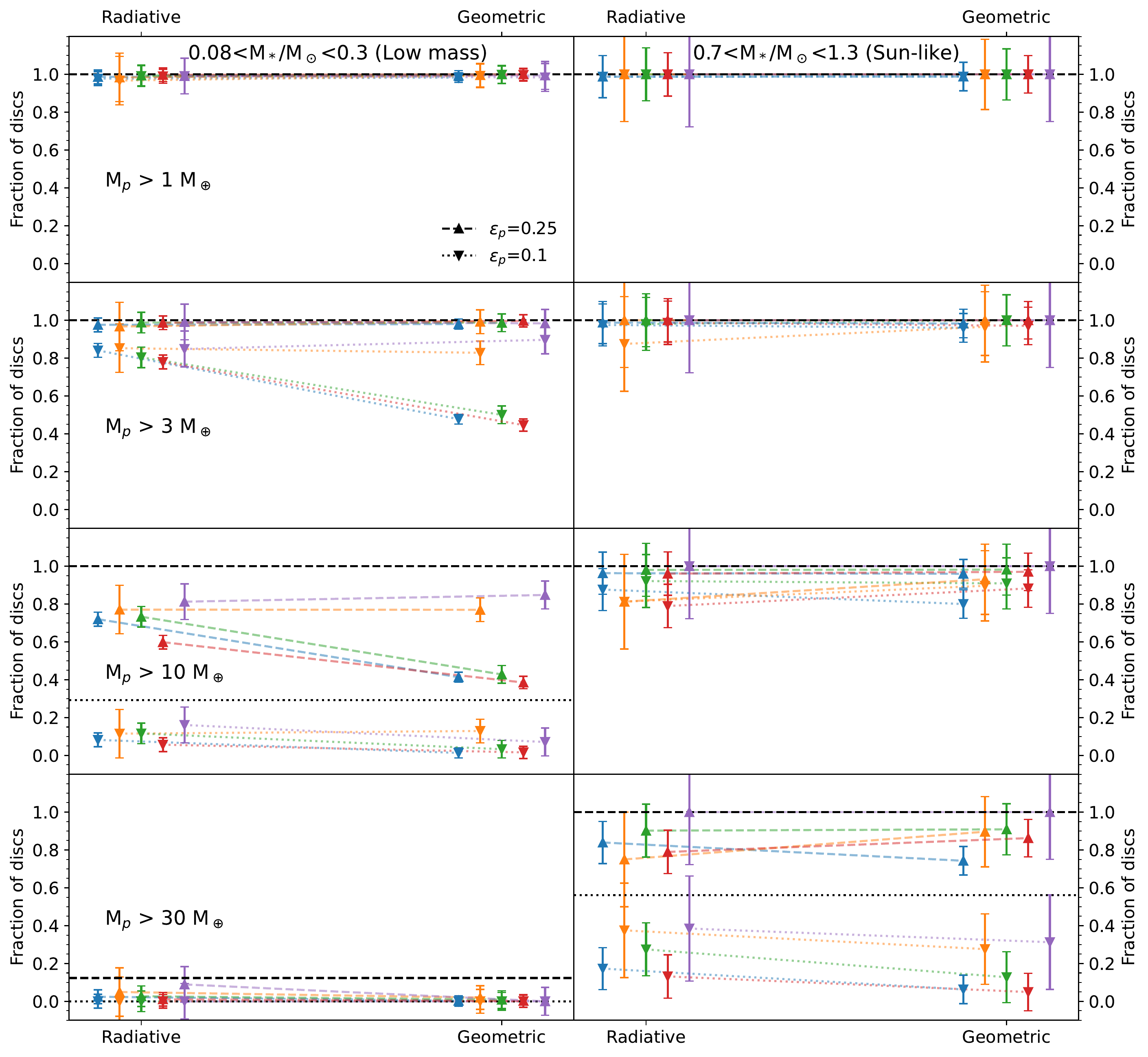}
    \caption{The fraction of discs with enough material, after dust EPE finishes, to form planetary systems of a given solid mass (top to bottom, 1, 3, 10, 30 M$_\oplus$), at a given planet formation efficiency. Discs are split into two samples based on their host star mass: 0.08-0.3 M$_\odot$ (low mass M dwarfs, left column) and 0.7-1.3 M$_\odot$ (Sun-like, right column). Lines are drawn between the data of the radiative and geometric runs of the same IMF realisation to guide the eye. As an example, in run m1r, about 85\% of discs around low mass stars has enough material to form 3 M$_\oplus$ of solid planets (at a formation efficiency of 10\%), while in run m1g this fraction is 50\%. Error bars denote one standard error of the Poissonian distribution for the total number of discs in each sample. Horizontal black lines show the fraction of discs with enough solid material at formation (for example, no discs around low mass stars exist that form with enough solid material to build 30 M$_\oplus$ of solid planets assuming 10\% planet formation efficiency); their line style is consistent with the corresponding planetary system mass.}
    \label{fig:planet_potential}
\end{figure*}

In Figure \ref{fig:planet_potential} we analyse the planet-forming potential of our disc populations by comparing their solid masses to certain mass limits. Given an amount of solid mass in planets, and a planet formation efficiency, we compute the fraction of discs that can form such planets. Note that we only consider the solid, or rocky/icy, component of the mass, not the gas component. As limits we choose 1, 3, 10, and 30 M$_\oplus$. These numbers are approximately logarithmically distributed and roughly correspond to a single terrestrial planet, a planetary system with multiple terrestrial planets or a super-earth, the core mass required to initiate runaway gas accretion \citep{Pollack1996}, and a rich, Solar system-like planetary system of terrestrial planets and gas giants. We also assume two values for the planet formation efficiency, 10\% and 25\%. These values were also used by \citet{Haworth2018a} when they introduced the dust EPE model to obtain constraints on the birth environment of the TRAPPIST-1 planetary system. They are also similar to the efficiencies required to form the observed giant planet core distribution from the dust component of Class 0 and Class I discs \citep{Tychoniec2020}.

In many cases the fraction of discs that can potentially form certain planetary (solid) masses is reduced from that when considering no loss of solid material. This is less important for lower planetary masses, higher efficiencies, and more massive host stars. For low mass stars, EPE can prevent the formation of one or more gas giant cores, and some super-earths or multiple terrestrial planets, especially in the absence of shielding. For Sun-like stars, EPE can prevent the formation of Solar system-like planetary systems, and some gas giant cores. Less massive planetary systems are barely affected.

Comparing the fraction of discs with enough solid mass for certain mass limits in paired runs, there are few cases where the difference is greater than one Poissonian standard deviation. The only cases where the difference is greater than one standard deviation (and in fact greater than $\sim$3 standard deviations) are runs m1r/g, m3r/g, and m4r/g, for low mass stars, at a planetary solid mass of 3 M$_\oplus$ and a formation efficiency of 0.1, and at a mass of 10 M$_\oplus$ and an efficiency of 0.25. For Sun-like stars, there is no significant decrease. Part of this is the larger standard deviation in the sample of Sun-like stars, but compared to the low mass sample there does not appear to be as large an absolute difference in fraction between paired runs. Runs m2r and m2g, and m5r and m5g, do not follow these trends, but these runs only formed a massive star in the main subcluster late in time (the first massive star in runs m5r/g formed in a smaller subcluster, and in runs m2r/g the first massive star formed 0.2 Myr later than in the other runs). As a result more discs are able to grow a large EPE-resistant solid reservoir before EPE begins. 

These results imply that radiation shielding can aid the retention of solid mass in discs around low mass stars. These discs can then potentially form a larger number of planets, or planets that are more massive. This effect is smaller, and not significant, for Sun-like stars. This is most likely due to lower EPE rates because their discs are more strongly bound.

\section{Discussion} \label{sec:discussion}

We have run simulations of protoplanetary discs in a star forming region. These simulations followed the collapse of a molecular cloud; the formation of stars with discs; stellar dynamics, feedback, and evolution; and disc evolution under environmental influences. We found that extinction decreases mass loss through external photoevaporation, increasing disc lifetimes and the reservoir of solids available for planet formation.

\subsection{Impact of the star formation algorithm on dynamic truncations}

The close encounters that result in dynamic truncations happen at length scales smaller than the size of star forming sinks. Additionally, the stars' initial velocities have a random component. As a result we under-resolve the early stellar dynamics at scales relevant for dynamic truncations. Properly resolving these dynamics requires higher resolution, in which every sink represents an individual protostellar core rather than a small region where stars form, such as in the simulations conducted by \citet{Bate2012,Grudic2021}\footnote{This does not yet account for binaries, which can have discs around each component and a circumbinary disc.}. However, this requires a major rework of our method and is outside the scope of this work.

The initial velocity structure will typically be erased on a scale of the free-fall time of the sink. In our main runs, we use a sink formation density of $1.528\cdot 10^{-20}$ g cm$^{-3}$, which yields a timescale of at most 0.5 Myr. A stellar population of mass $M_c$ within the sink radius $r_s=0.17$ pc will have a free-fall time of $1.2 \left(M_c/M_\odot\right)^{-0.5}$ Myr. Even with a stellar population of $\sim$1000 M$_\odot$ (comparable to the total mass of stars formed) within a sink, that only reduces the free-fall time to $\sim$0.04 Myr, still a considerable fraction ($>$1\%) of our simulation time and longer than the time resolution of most of the figures presented. As a result the initial velocity structure will leave an imprint for a non-negligible amount of time.

\begin{figure}
    \centering
    \includegraphics[width=1\linewidth]{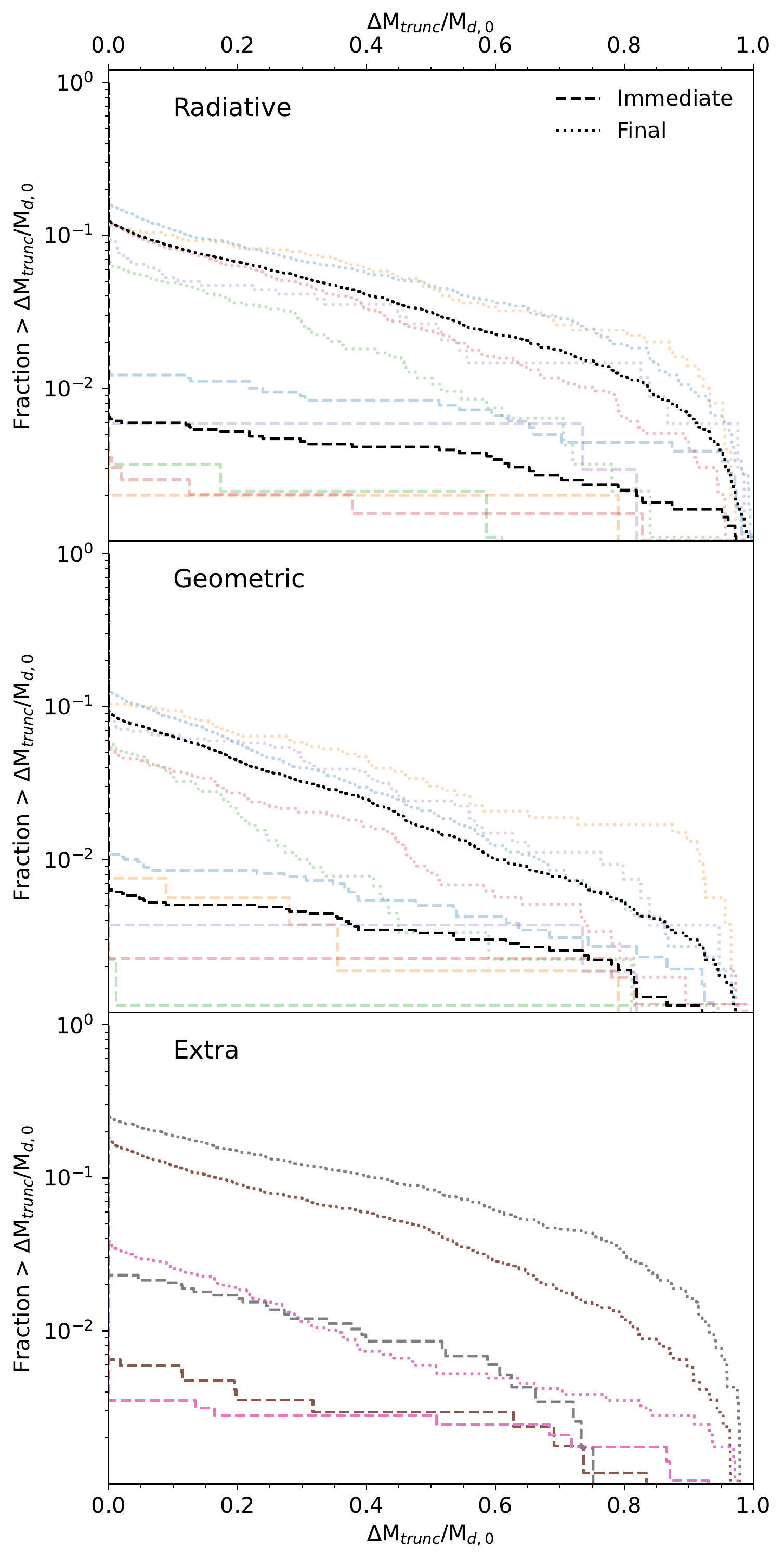}
    \caption{The fraction of discs that have lost mass due to dynamic truncation greater than some value, within the first few kiloyears after formation (dashed) and at the end of the simulation (dotted). The top panel shows main radiative runs, the middle panel the main geometric runs, and the bottom panel the extra runs. Black lines are aggregated over all simulations in each frame.}
    \label{fig:trnc_im_end}
\end{figure}

Additionally, the stellar density at formation will be determined by the resolution when the interval between two formations is shorter than the time it takes to leave a sink particle. Because the density influences the distribution of encounter distances, we expect the effect of dynamic truncations to depend on resolution. The probability that a newly formed star is initialised on an orbit leading to a nearly instantaneously truncation event also increases at higher resolution. In general, we find that most truncation events happen within the volumes of sinks, but involve stars that have already left their birth sink. In Fig. \ref{fig:trnc_im_end}, we show the distribution of truncation mass loss shortly (up to 10 kyr) after each star's formation and at the end of each simulation.

A fraction of $\sim$0.3\%-1\% of all discs in medium-resolution runs experience a truncation event shortly after formation (this increases to $\sim$3\% in the high resolution run). These truncation events can lead to the loss of $\sim$95\% of the initial disc mass. However, many more truncation events happen at a later time, to the point that 3\%-30\% of discs have undergone a truncation. Thus, early truncations affect a small but non-negligible fraction of all discs, but the bulk of mass loss through truncations happens later.

Resolution affects the importance of dynamic truncations. Both the immediate and final mass loss is greater in run h3r than in l3r, except for a small tail at high immediate mass loss. The final mass loss in l3r and h3r are also respectively smaller and greater than in m1-5r. Note that run h3r has not advanced as far in time as l3r and m3r, so the existing discs could undergo even more truncation events. New stars will also form, but because of the (for now) exponential star formation rate, the density of newly formed stars within sinks will only increase, likely increasing the number of immediate and total truncations.

We conclude that the star formation algorithm used in Torch affects the results of dynamic truncations. To properly resolve this we'd need to increase resolution such that every sink particle is its own star. However, our general findings that EPE is globally dominant over dynamic truncations \citep[consistent with ][]{Winter2018}, but that dynamic truncation can dominate in individual discs, hold across a range of resolutions.

\subsection{Pre-main sequence evolution}

We assume that stars are born on the zero-age main sequence (ZAMS). We thus neglect the embedded phase of disc evolution, and we neglect the pre-main sequence evolution of the feedback strength of massive stars. This includes FUV luminosity, so this also affects EPE.

\subsubsection{Embedded discs}

During the first 0.5 Myr of the formation of low-mass stars \citep{Dunham2014}, the star and its disc are embedded in an envelope. This envelope is another source of extinction of the FUV radiation field and will further shield the disc from EPE. This can resolve the issue of our short disc lifetimes as compared to observations. However, this envelope will also itself be subject to EPE, which may shorten the embedded phase (and even impact the formation of the star and disc).

\subsubsection{Pre-main sequence massive stars}

The suddenness with which EPE mass loss accelerates when the first massive star forms is an artefact of the identically zero FUV luminosity of lower mass stars and the instantaneous formation of massive stars with non-zero FUV luminosity. 

\begin{figure}
    \centering
    \includegraphics[width=1\linewidth]{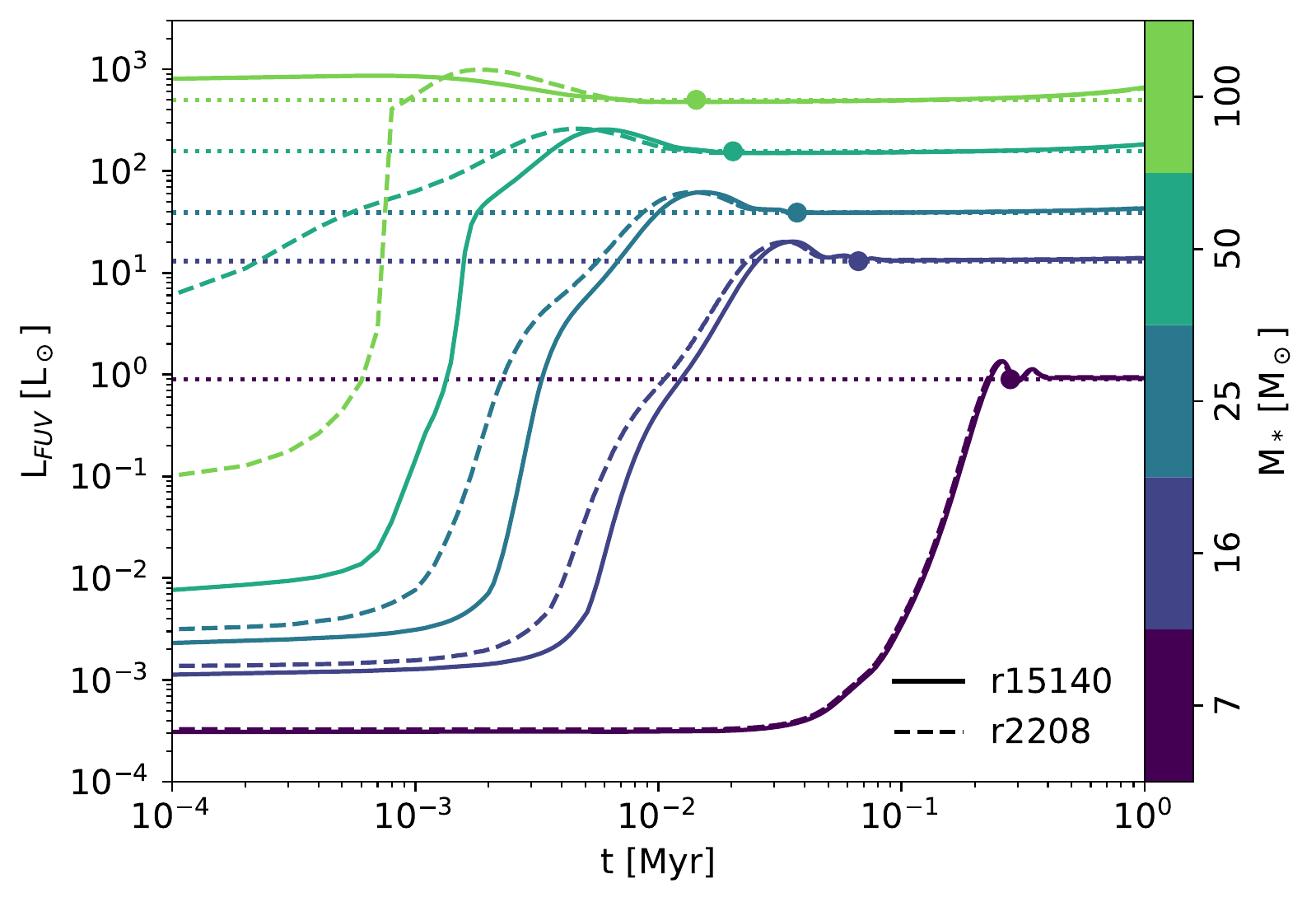}
    \caption{The evolution of the FUV luminosity of massive pre-main sequence stars. The dotted horizontal lines denote the ZAMS FUV luminosity. Points indicate the ZAMS Kelvin-Helmholtz timescale. Results are obtained using MESA releases 2208 (dashed lines) and 15140 (solid lines).}
    \label{fig:pre_ms}
\end{figure}

In Fig. \ref{fig:pre_ms}, we show the evolution of the FUV luminosity of a selection of massive pre-main sequence stars. To obtain these values, we set up pre-main sequence stellar models in the MESA stellar evolution code \citep[releases 2208 and 15140;][]{Paxton2011,Paxton2019}, integrated their black-body spectra over the FUV spectral range, and multiplied those with the stellar surface area (the same procedure used in our model). These stellar models are initialised with a metallicity of 0.02. For comparison, we also plot the FUV luminosity of ZAMS from our model for the same masses. We also indicate the ZAMS Kelvin-Helmholtz timescale of each star.

The FUV luminosity of pre-main sequence stars is initially lower than the ZAMS value for $<1$ kyr for a 100 M$_\odot$ star to 0.2 Myr for a 7 M$_\odot$ star (all very close to the ZAMS Kelvin-Helmholtz timescale). There is also a period of 10-100 kyr where the luminosity is up to a factor $\sim$2 larger than the luminosity at the ZAMS. These time scales are a considerable fraction of our run times. The results of the two versions of MESA for the pre-main sequence are similar for lower mass, but differ considerably at early times for higher masses. The main sequence evolution is nearly identical across stellar mass. 

In our simulations, most massive stars are on the lower end of the mass range. The radiation field surrounding these stars should then be built up more gradually, resulting in a more gradual increase of EPE mass loss. We argue that the pre-main sequence phase with negligible luminosity is represented by the sink particle while it is accreting sufficient mass. However, there is typically a 10-100 kyr period where the luminosity is within a factor 10 of the ZAMS luminosity, which is a considerable fraction of our simulation time.

The number of massive stars in our simulations is $\sim$10, which would make it feasible to evolve those stars using MESA. Our simulations only cover a relatively short timescale, during which the stars remain on the main sequence. Alternatively, pre-computed pre-main sequence tracks could be used, such as the MIST tracks \citep{Choi2016}.

\subsection{Variance of star-forming regions}

Our simulations represent star-forming regions that are relatively massive, like the Orion Nebula Cluster, as compared to the regions in the Solar neighbourhood where we are able to properly study protoplanetary discs. Other regions such as Lupus or Taurus contain fewer stars and lack OB stars. EPE in these regions may be driven by less massive stars (which we neglected), or even by massive stars from a nearby, more massive star-forming region. This is thought to happen in Lupus, where the disc IM Lup is irradiated by massive stars from the Galactic field and nearby young associations, and the radiation field is $\sim$4 G$_0$ \citep{Cleeves2016}. This also implies that for low mass star-forming regions, we need EPE rates for radiation fields $<10$ G$_0$, which are currently not available in the FRIED grid. 

We are currently unable to properly simulate protoplanetary discs in low mass star forming regions because protostellar outflows are not implemented in our model, and \citet{Matzner2000} pointed out that cloud disruption is dominated by protostellar outflows when the initial cloud mass is $<10^{3.5}$ M$_\odot$.

\subsection{Comparison with earlier simulations}

\subsubsection{Correlations of disc mass with distance to massive stars}

\citet{Parker2021b} investigated the correlation between disc mass and projected distance to a massive star that can be induced by EPE. Such a correlation is found in multiple star forming regions, such as the ONC \citep{Mann2014,Eisner2018} and $\sigma$ Orionis \citep{Ansdell2017}. \citet{vanTerwisga2019} found that within the Orion Molecular Cloud, discs in the Trapezium cluster (which hosts an O-type star) are less massive than those outside it. \citet{Parker2021b} suggest that such a correlation can be polluted by projection effects (a star close in projection being a back- or foreground cluster star) and cluster dynamics (close, much-evaporated discs moving further away). They find significant correlations between disc mass and projected distance to the most massive star in only a minority of their simulations, and conclude that the observed correlation is likely not caused by EPE.

However, their simulated clusters contain multiple massive stars, and the most massive star is frequently only slightly more massive than the second most massive star. In such clusters the radiation field may not everywhere be dominated by the most massive star. A population of evaporated discs close to the second most massive star may remove the correlation of disc mass with distance to the most massive star. The signature of EPE may then be found in the correlation between disc mass and the distance to the star that dominates its local radiation field. 

Our models take into account processes that can further complicate this problem, of which we briefly discuss the impact on the correlation. Ongoing star formation leads to a spread in stellar ages. The younger population will typically have more massive discs, and may inhabit a different region of space than the older population. We may expect that the younger population is closer to the cluster centre. This is also where the massive stars migrate to due to mass segregation. This would create a negative correlation between mass and projected distance, rather than a positive correlation as observed. Extinction, on the other hand, can be expected to contribute to a positive correlation by shielding further-away discs. 

The disc mass tracer also complicates whether a correlation can be established. The disc masses used to establish the observed correlation were based on observations of the dust component. Although a dust-to-gas ratio of 0.01 is typically used to estimate the total disc mass, this ratio can be altered by, e.g., pebbles drifting onto the star, solids being locked in planetesimals and/or planets, and EPE. For the correlation, this ratio is only important if it is affected by EPE. Dust EPE only reduces the dust mass in young discs, while the grains are small enough to be entrained in the wind. The net effect is a population of discs not affected by EPE, making the correlation less strong and significant. Finally, \citet{Haworth2021b} pointed out that the radiation responsible for EPE can also heat a disc's dust, which leads to an overestimate of the dust mass\footnote{The observed radiation can be supplied by a smaller amount of higher temperature, more luminous dust.}. This effect would make a positive correlation appear less strong.

Finally, we note that ongoing star formation results in differences in cluster structure and evolution. Of the two cluster models explored by \citet{Parker2021b}, the high density model has a stellar density roughly comparable to our simulation. They find no considerable differences in the correlations between high and low density clusters.

We perform an analysis of the mass-distance correlation for our data. Following \citet{Parker2021b} we use the Spearman rank coefficient test to determine the strength and significance of the correlation between disc mass (both gas and solid) and projected distance to the most massive star. We also perform the same analysis with the distance to the nearest massive star. We follow the evolution of the correlation coefficient and the p-value\footnote{Of the null hypothesis that the data are uncorrelated, i.e. the probability that a random realisation has an equal or stronger correlation.} through time. For every moment in time, we perform the analysis 300 times under different random rotations, in order to reduce biases due to projection effects. 

\begin{figure*}
    \centering
    \includegraphics[width=1\linewidth]{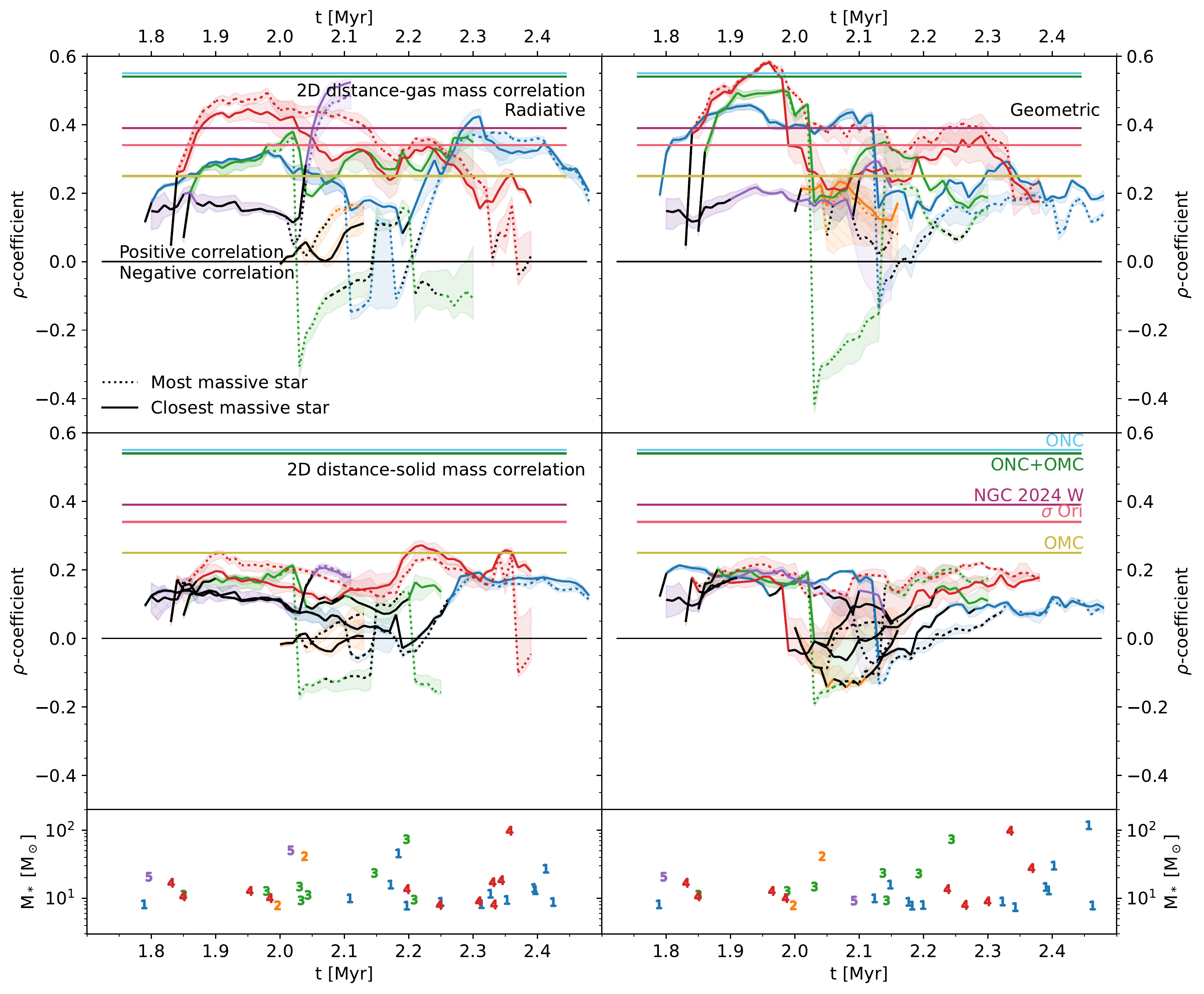}
    \caption{The time evolution of the Spearman's rank correlation coefficient of the correlation between disc gas (top panel) and solid (middle panel) mass and 2D projected distance to the closest (solid central line) and most massive (dotted central line) of the massive stars, in the main runs. The bottom panels show the moments of formation and masses of massive stars. The left column shows the radiative runs, the right column the geometric runs. These results are obtained from projection after 300 random rotations. Shaded regions denote the $16^\textrm{th}$ and $84^\textrm{th}$ percentiles, the lines the median. The line is a run's colour if the correlation is significant (median p-value $<0.01$), black if not significant. Coloured horizontal lines indicate significant correlation coefficients reported by \citet{Parker2021b} for the correlation between 2D projected distance to the most massive star and disc dust mass in the Orion Nebula Cluster (ONC), Orion Molecular Cloud, ONC and OMC combined, and $\sigma$ Ori. The value for NGC 2024 W was obtained by \citet{Haworth2021a}.}
    \label{fig:dist_corr_compact}
\end{figure*}

In Fig. \ref{fig:dist_corr_compact} we show the evolution of the Spearman's rank correlation coefficient of the correlation between disc gas or solid mass and 2D projected distance to the closest massive star or most massive star, for our main runs. For reference, we show the formation times and masses of the massive stars in these runs. We also show, for a number of star forming regions, the observed coefficients for the (p-value $<$0.01) correlation between solid mass and 2D projected distance to the most massive star. 

For brevity, we refer to the correlation between gas mass and distance to the closest massive star as c-gCMS, the correlation between solid mass and distance to the most massive star as c-sMMS, with c-sCMS and c-gMMS defined analogously.

In general we find significant positive c-gCMS. Except for run m2r and run m5r (until the formation of the second massive star, which is the first in the large subcluster) all runs quickly establish a significant correlation. This is occasionally broken (run m1r at 2.2 Myr, run m5g at 2.1 Myr), but these moments correspond to the formation of a massive star. If a massive star forms in a cluster of stars that has seen little irradiation thus far, it counts as the closest massive star but has not yet been able to photoevaporate nearby discs. However, the c-gCMS is then quickly re-established. Between radiative and geometric runs there are variations in the correlation coefficient, but there is no clear trend in which of these types of runs has stronger correlations. 

The c-gMMS frequently follows that of the correlation with the distance to the closest massive star. If only one massive star is present c-gCMS and c-gMMS are identical by definition, but even after the formation of a second massive star they are frequently similar. At some moments c-gCMS and c-gMMS diverge considerably, e.g. in runs m1r/g at 2.1 Myr and runs m3r/g at 2.0 Myr, where there is even a significant negative correlation for a time (plus periods of non-significance). These moments correspond to the formation of a star more massive than any formed prior. At these moments, the coefficient of the correlation with distance to the closest massive star also becomes smaller, though the correlation remains significant. This implies that EPE is indeed not dominated by the most massive star everywhere in the cluster, and the non-significant correlations found by \citet{Parker2021b} -- and the accompanying conclusion that external photoevaporation is not responsible for the observed correlations -- may be due to not accounting for other massive stars.

The c-sCMS and c-sMMS are typically weaker than c-gCMS and c-gMMS. The correlation is also more frequently not significant. This is due to the decrease in EPE of dust due to dust growth. An old disc with a large remaining solid reservoir that comes close to a massive star loses less dust than gas.

The observed correlation coefficients of the OMC, $\sigma$ Ori, and NGC 2024 W are in good agreement with the typical coefficients for the correlations with gas mass. Those of the ONC and ONC+OMC are on the high side, with only run m4g briefly reaching a similarly strong correlation. The coefficients for the correlation with solid mass are lower than the observed coefficients; only m4r briefly achieves a correlation comparable to the OMC's, but not any other region. While neither tracer is truly representative of the tracer used in observation, the observed dust mass should be closer to the solid mass than to a scaled gas mass.

Interestingly, the observed correlations may even be underestimates, because they do not account for the overestimated mass of irradiated dust \citep{Haworth2021a}, further increasing the discrepancy with our model. If our model of dust EPE is correct, there may indeed be an additional process that leads to the observed correlations. \citet{Sellek2020} introduced a more detailed model of dust EPE that included drift. They find that generally, dust that is large enough to not be entrained is large enough to drift inwards. In practice, this has a similar effect to our assumption that the gas-to-dust ratio at the outer disc edge is always 1\%: large dust is not entrained in either case.

\citet{Parker2021b} suggest dynamic truncations as an alternative, but this is a process we also take into account. Accounting for dust drift would make truncations even more ineffective, because the more compact dust distribution would be more resistant to truncation than the gas distribution.

\subsubsection{Protoplanetary discs in a different star forming region}

\citet{Qiao2022} use a pre-computed model of a star forming region to inform the radiation field to which their disc population is exposed. Their radiative transfer includes extinction by cluster gas. They do not resolve the dynamics of individual stars (neglecting dynamic truncations), but group them together in sink particles. Inside these sink particles the distance between a radiation source and a disc is assumed to be half a cell size, or 0.09 pc. The initial cloud they consider is more massive than ours, $10^5$ M$_\odot$ as opposed to $10^4$ M$_\odot$. 

The time for discs to be exposed to an unshielded radiation field is $\sim$0.5 Myr in their simulations. Contrast this with our simulations, where 0.5 Myr after the formation of the first massive star (and as late as 0.8 Myr, in runs that reach that far) the radiation field is still extincted for the majority of discs.

We note that the initial virial ratio of our cloud differs considerably from that of \citet{Qiao2022}'s. They use a supervirial, marginally bound cloud ($\alpha_v=2$), while our cloud is strongly bound ($\alpha_v=0.25$). \citet{Dale2012,Dale2013} discussed how ionising feedback can clear feedback bubbles more easily in initially unbound clouds, compared to clouds with lower virial ratios. Our simulations do not show multiple parsec scale feedback bubbles, likely due to a combination of lower virial ratio and a shorter period of feedback ($\sim$1 Myr, compared to $\sim$3 Myr in the work of \citet{Dale2012,Dale2013}).

Surveys of molecular clouds typically find median virial ratios greater than unity for clouds with masses $\sim 10^4$ M$_\odot$ \citep{Chevance2022}, though with a typical spread of one order of magnitude \citep{Evans2021}. This argues for shorter shielding times than we find. Still, this implies that the evolution of protoplanetary disc (and as a consequence the formation of planets) can depend not only on the mass of a cloud, but also its dynamical state. 

An upper limit on the gas clearing time has been derived by \citet{Kim2022}. By 3 Myr, the emission of molecular gas and young stars in nearby galaxies have become spatially decoherent on length scales $\sim$100 pc. This leaves a range of gas clearing times of 0.5-3 Myr. Whether EPE can impact planet formation in a majority of discs remains uncertain, depending on how quickly planets form.

\subsection{Comparison with observations of NGC 2024}

\citet{vanTerwisga2020,Haworth2021a} present observations of proplyds in NGC 2024 (the Flame Nebula). This region contains two spatially separated subpopulations of stars. \citet{vanTerwisga2020} find that the more embedded, $\sim$0.2--0.5 Myr old population has a disc fraction of 45\%, and the less embedded, $\sim$1 Myr old population has a disc fraction of 15\%. Both subpopulations also contain a massive star (an O8V star in the younger region, and a B0.5V star in the older region). \citet{Haworth2021a} find proplyds in both subpopulations, whose EPE appears to be dominated by the massive star in their subpopulations, without apparent influence of the massive star in the other. This is similar to the results of \citet{Odell2017}, who found that a number of proplyds in the ONC are being photoevaporated by $\theta^2$ Ori A, the cluster's second most massive star, rather than $\theta^1$ Ori C, as commonly assumed. Both of these findings agree with our earlier finding that less massive stars may dominate their local EPE.

\citet{Haworth2021a} also perform rank correlation tests on the disc mass (as derived from emission of dust) and projected distance to the massive stars. When splitting the discs into subsets depending on the closest massive star, they find a significant correlation for the stars closer to the older massive star (Spearman 0.39, p-value $2\times 10^{-4}$; we include this value in Fig. \ref{fig:dist_corr_compact}), but not for the stars closer to the younger massive star. Considering the full population, there are no significant correlations with the distance to either massive star. This is in agreement with our suggestion that the effects of EPE on the distribution of disc masses throughout the cluster can be obscured if only the distance to the most massive star is considered.

The one significant correlation coefficient is much higher than any value we find for the correlation with solid mass, although it would be a reasonable value for correlation with gas mass. Our results show that a significant correlation with gas mass can be re-established in $\ll$0.1 Myr, much shorter than the age of the young region. In the correlation with solid mass there are cases were a significant correlation is re-established in $\sim$0.2 Myr (m3r, m1g, m4g), but the formation of multiple massive stars in these intervals complicates this reasoning. 

The relatively low disc fraction in NGC 2024 is an interesting contrast to our results. Our simulations cover a time interval longer than the age of the younger region, but $>$99\% of stars with an initial disc retain it. Including stars more massive than 1.9 M$_\odot$ lowers this disc fraction, but only 4\% on the Kroupa IMF are more massive than that. This can't lower the fraction to the 45\% observed in the younger subpopulation of NGC 2024. This could point towards our initial discs being too massive, at least on average. \citet{Kuffmeier2020} point out that increased cosmic-ray ionisation rates in protostars near high mass stars can lead to the formation of smaller discs. However, this does not affect discs formed before the first massive star.

\section{Summary \& conclusion} \label{sec:conclusion}

We have performed simulations that coupled a model of a star forming region (including hydrodynamics and stellar dynamics, evolution, and feedback, with the interactions resolved in the Astrophysical MUltipurpose Software Environment) with a model of a population of protoplanetary discs. With this coupled model we studied the effects that the environment of the star forming region has on the protoplanetary discs through external photoevaporation and dynamic truncation. We focused on the effects of radiation extinction due to remaining gas of the star forming cloud which shielded protoplanetary discs from external photoevaporation. We ran initially identical simulations with and without extinction.

\begin{itemize}
    \item The evolution of a young star cluster, and its population of protoplanetary discs, heavily depend on where and when massive stars form, and how massive they are. This variation is likely larger than that introduced by differences in the orientation of the cloud's initial velocity field.
    \item Cluster gas left over from star formation shields the majority of discs for at least 0.5 Myr after the formation of the first stars that drive both external photoevaporation and feedback. This shielding results in less external photoevaporation throughout the disc population. Even discs that experience strong radiation fields do so for shorter amounts of time. 
    \item The lower limit of 0.5 Myr for the duration of shielding contrasts shorter durations found in earlier work. We attribute this discrepancy to the low virial ratio of our initial cloud, which is at the lower extreme of observed molecular clouds. This implies that the importance of external photoevaporation for the evolution of protoplanetary discs (and the formation of planets) depends on the dynamical state the discs form in.
    \item Runs with extinction had more mass loss through dynamic truncations than runs without extinction. This reveals a competition between EPE and dynamic truncation, which both operate at the outer disc edge. We also find that EPE dominates mass loss in the overall disc population, but that in up to 10\% of discs dynamic truncations dominate mass loss over EPE. Most dynamic truncations in our simulations are to radii smaller than those observed in star forming regions, but our distribution is skewed towards smaller radii due to EPE shrinking discs throughout the population.
    \item Although there is a trade-off between EPE and dynamic truncation, the net effect of lowering EPE through shielding is less mass loss, and longer disc lifetimes by up to an order of magnitude. This slightly alleviates the problem of very short disc lifetimes in earlier simulations of EPE in clusters. We also find that in star forming regions with massive stars, discs around higher mass stars have longer lifetimes than those around lower mass stars. This is in line with earlier simulations of EPE in young stellar populations, but disagrees with observations. Notably, this is with the inclusion of factors that could alleviate this discrepancy (decreasing disc mass-stellar mass ratio with increasing stellar mass, and IPE mass loss that is super-linear in stellar mass). 
    \item The extraction of dust from discs through EPE can reduce the planet forming potential of discs. For our initial disc masses, radiation shielding can increase the fraction of low mass stars that have the solid mass to form systems of terrestrial planets and gas giant cores. For Solar-like stars, this effect is not significant. 
    \item We recover consistently significant correlations between disc gas mass and projected distance to the closest massive star. The strengths of these correlations are broadly consistent with those of observed correlations between disc solid mass and projected distance. Correlations with the disc solid mass from our models are less strong than observed, and not consistently significant. We may overestimate the resistance of dust to external photoevaporation, or there may be a process at work here that we have not accounted for. We also find that considering only the most massive star in such analyses can lead to non-significant correlations, and stress that external photoevaporation can be dominated by different stars in different parts of a cluster. The signature of EPE on the distribution of disc masses throughout a cluster must then be sought by considering the stars that dominate each disc's EPE.
\end{itemize}

We have shown that radiation shielding by gas in massive star forming regions strongly affects the evolution of protoplanetary discs within the cluster. It increases disc lifetime by reducing external photoevaporation, giving more time for planets to form. It also leaves discs around low mass stars with more solid mass, allowing for the formation of more massive planetary systems.

\section*{Acknowledgements}

We would like to thank the referee for their fast and insightful comments, which have helped improve the quality of this paper.

The simulations in this work have been carried out on the Snellius supercomputer, hosted by the Dutch national high performance computing center SURF. About 1.2 million CPU hours have been used for the simulations. Additional CPU hours have been used for tests, reruns, and analysis, putting the total at $\sim$2 million. The AMD Rome 7H12 CPUs use 280 W, with 64 cores each, putting the total power usage at 8.75 MWh, about two to three times the yearly electricity usage of an average household. SURF runs fully on renewable energy.

MJCW is supported by NOVA under project number 10.2.5.12. CCC is supported by a Canada Graduate Scholarship -- Doctoral from the Natural Sciences and Engineering Research Council of Canada. AT was partly supported by NASA FINESST 80NSSC21K1383, and a NASA Cooperative Agreement awarded to the New York Space Grant Consortium. AT and M-MML acknowledge partial funding from NSF grant AST18-15461.

We would like to thank the rest of the Torch team for interesting discussions and a pleasant collaboration, including Sabrina Appel, Will Farner, Joe Glaser, Ralf Klessen, Stephen McMillan, and Alison Sills. We also want to thank the AMUSE team for their continuing development of the framework, especially Inti Pelupessy and Steven Rieder.

\section*{Data Availability}

The data underlying this article will be shared on reasonable request to the corresponding author.

The code underlying this article is available at \url{https://bitbucket.org/torch-sf/torch/src/main/}\footnote{Git commit used: 7c90f85677e8a3b9316cbd4eaaada515737af08e}, \url{https://github.com/MJCWilhelm/ppd-population}\footnote{Git commit used: d80f6e17d86cef8405a3fb990e10f5a47e28d7bc}, and \url{https://github.com/amusecode/amuse}\footnote{Git commit used: ffe1b47b8077bc9d991bfb5b3ce911ed7052edf8}.

\section*{Software}

Simulations were done using the AMUSE framework \citep{PortegiesZwart2009,Pelupessy2013}, using the Torch model \citep{Wall2019,Wall2020}. This model couples ph4 \citep{McMillan2012}, SeBa \citep{PortegiesZwart1996,Toonen2012}, and FLASH \citep{Fryxell2000} \citep[which is in turn coupled to FERVENT,][]{Baczynski2015}. Additionally, we coupled VADER \citep{Krumholz2015} to Torch, and used MESA \citep{Paxton2011,Paxton2019} for additional analysis. The data analysis for this paper was done in Python \citep{van1995python}, with the NumPy \citep{Harris2020} and SciPy \citep{Virtanen2020} modules. Plots were made using the Matplotlib \citep{Hunter2007} and yt \citep{Turk2011} modules.



\bibliographystyle{mnras}
\bibliography{references} 




\appendix

\section{Flux estimation in Torch} \label{app:flux}

The radiative transfer module of Torch, FERVENT, computes the FUV radiation field flux in a cell by summing up the energy flow of all rays intersecting the cell, and dividing that by the face area of the cell. However, the cross section (or equivalently, the shadow) of a cube is not the same at every angle of incidence. Only when the incidence vector is normal to a cube face is the cross section equal to the face area. As we will demonstrate, the cross section is larger than a cube face when viewed at a different angle. As a result, the original method typically underestimates the cross section, and then overestimates the radiation field. We introduce an approximate correction based on the cross section of a cube rotated around one axis normal to its faces. 

Given a cube viewed along a face-centred axis, the shadow is simply the area of a face. If the sides have length $d$, this is:

\begin{equation}
    A = d^2.
\end{equation}

\begin{figure}
    \centering
    \includegraphics[width=0.5\linewidth]{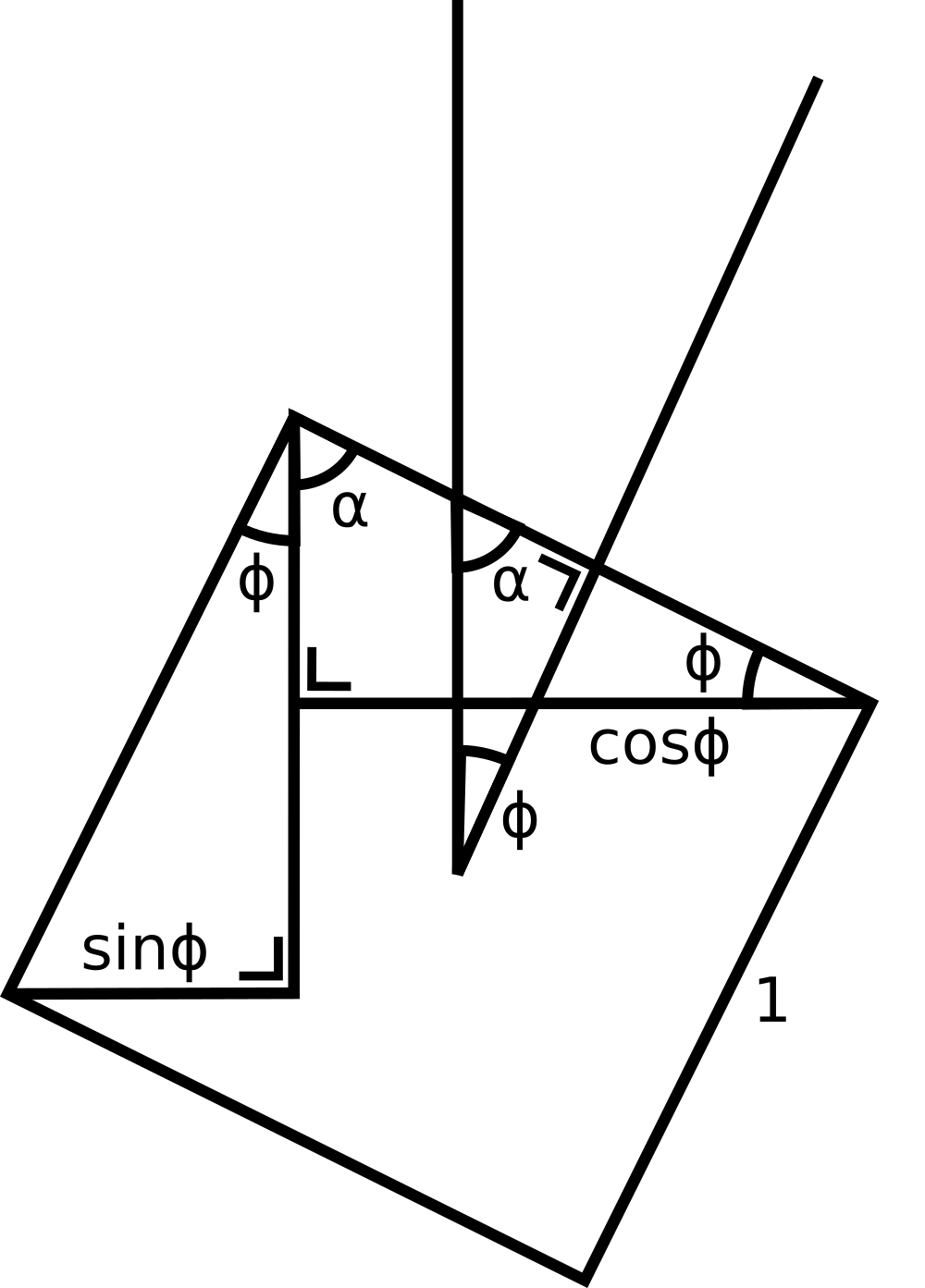}
    \caption{A unit cube rotated by an angle $\phi$. $\alpha$ is an auxiliary angle, and is $\pi/2-\phi$. The cross section when viewed at an angle $\phi$ is the sum of two components, $\sin\phi$ and $\cos\phi$.}
    \label{fig:rot_cube}
\end{figure}

Rotation around another face-centred axis by an angle $\phi$ shrinks the effective cross section of the formerly top face, but exposes another face (see Fig. \ref{fig:rot_cube}). The side parallel to the rotation axis will remain constant, but the projected sides of the two faces will change. Identifying triangles in Fig. \ref{fig:rot_cube}, we find the following expression:

\begin{equation}
    A\left(\phi\right) = d\cdot d\left(\sin\phi + \cos\phi\right).
\end{equation}

By symmetry, the projected area by rotation about the other axis ($A\left(0,\theta\right)$) has the same form. 

Our approximation directly combines these two limiting cases:

\begin{equation}
    \tilde{A}\left(\phi,\theta\right) = d^2\left(\sin\phi + \cos\phi\right)\left(\sin\theta + \cos\theta\right).
\end{equation}

\begin{figure}
    \centering
    \includegraphics[width=0.5\linewidth]{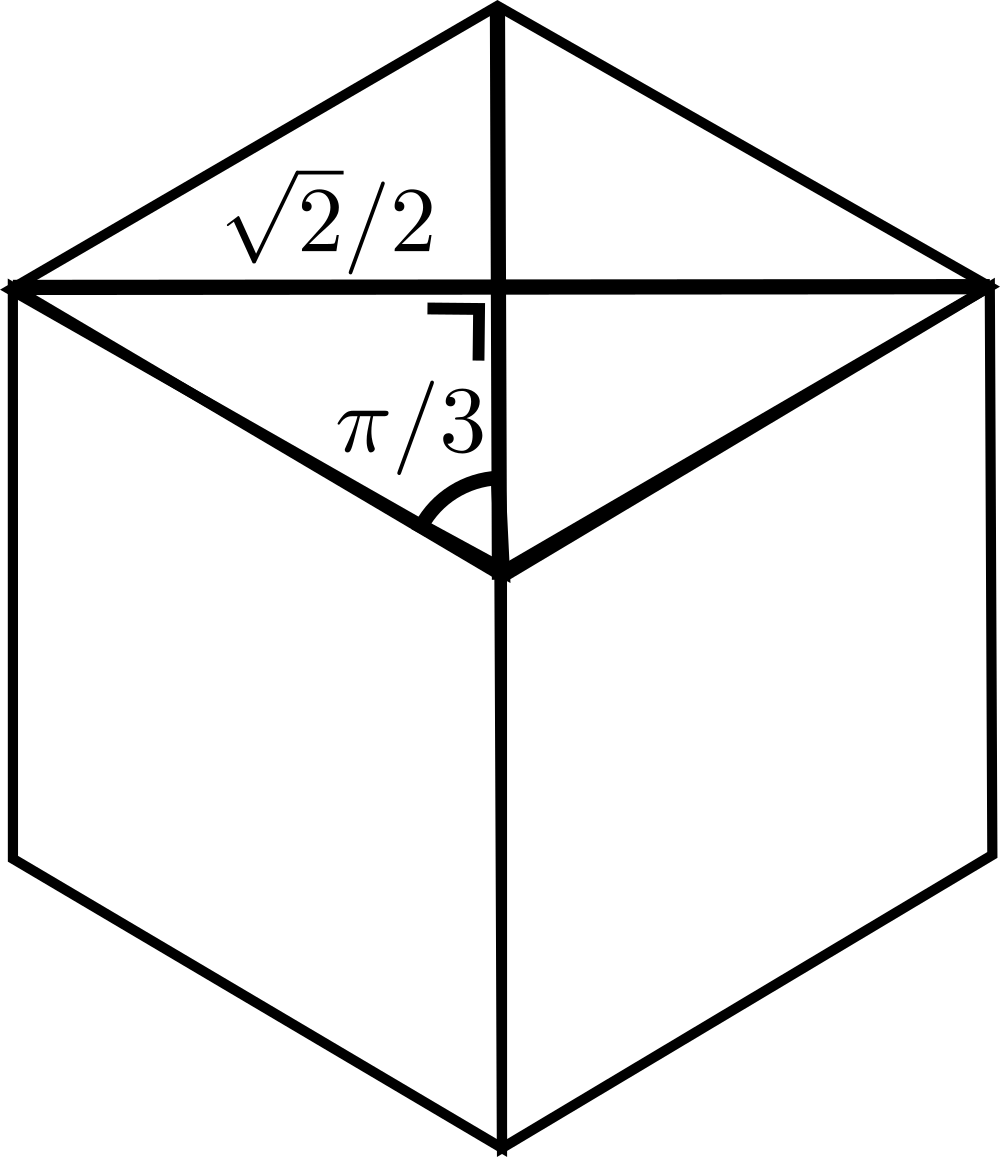}
    \caption{Top view of a unit cube, rotated by $\pi/4$ and $\arctan\sqrt{2}$ around the axes in the viewing plane. The first rotation extended the cross section's width to $\sqrt{2}$, and the second rotation left this invariant. Therefore, the long axis of the top rhombus is still $\sqrt{2}$. We divide this rhombus in four equal right triangles, which have one leg equal to $\sqrt{2}/2$ and the opposing angle $\pi/3$ (from a full $2\pi$ divided in six equal parts). The other leg then has length $\frac{\sqrt{2}/2}{\tan\pi/3}=1/\sqrt{6}$. The triangle's area is then $\left(\sqrt{2}/2\right)\left(1/\sqrt{6}\right)/2=1/\left(4\sqrt{3}\right)$. The full cross section consists of 12 such triangles, so the total area is $12/\left(4\sqrt{3}\right)=3/\sqrt{3}=\sqrt{3}$.}
    \label{fig:top_view}
\end{figure}

We can estimate the quality of this approximation by investigating a special case. Rotating one axis by an angle of $\pi/4$, and the other by an angle of $\arctan\sqrt{2}\approx 0.3\pi$, we view the cube along an axis passing through two extreme corners\footnote{Assuming a spherical coordinate system with $\hat{r}=\cos\phi\sin\theta\hat{x} + \sin\phi\sin\theta\hat{y} + \cos\theta\hat{z}$, setting $\phi=\pi/4$ and $\theta=\arctan\sqrt{2}$ yields a unit vector $\hat{r}=\left(\hat{x}+\hat{y}+\hat{z}\right)/\sqrt{3}$.}. Fig. \ref{fig:top_view} demonstrates how we can compute the effective cross section of this configuration: $A\left(\pi/4,\arctan\sqrt{2}\right)=\sqrt{3}d^2$. In our approximation, $\tilde{A}\left(\pi/4,\arctan\sqrt{2}\right)\approx 1.97d^2$. The relative error between the true value and the approximation is $\tilde{A}/A-1\approx0.14$.

We test our approximation in practice by setting up a simple test problem in FLASH and comparing both methods. We place a radiation source with luminosity $L$ in the origin, and set the global radiation cross section of the gas to 0. This test was run with a minimum and maximum refinement level of 2, but the results did not differ significantly at refinement level 3. 

\begin{figure}
    \centering
    \includegraphics[width=1\linewidth]{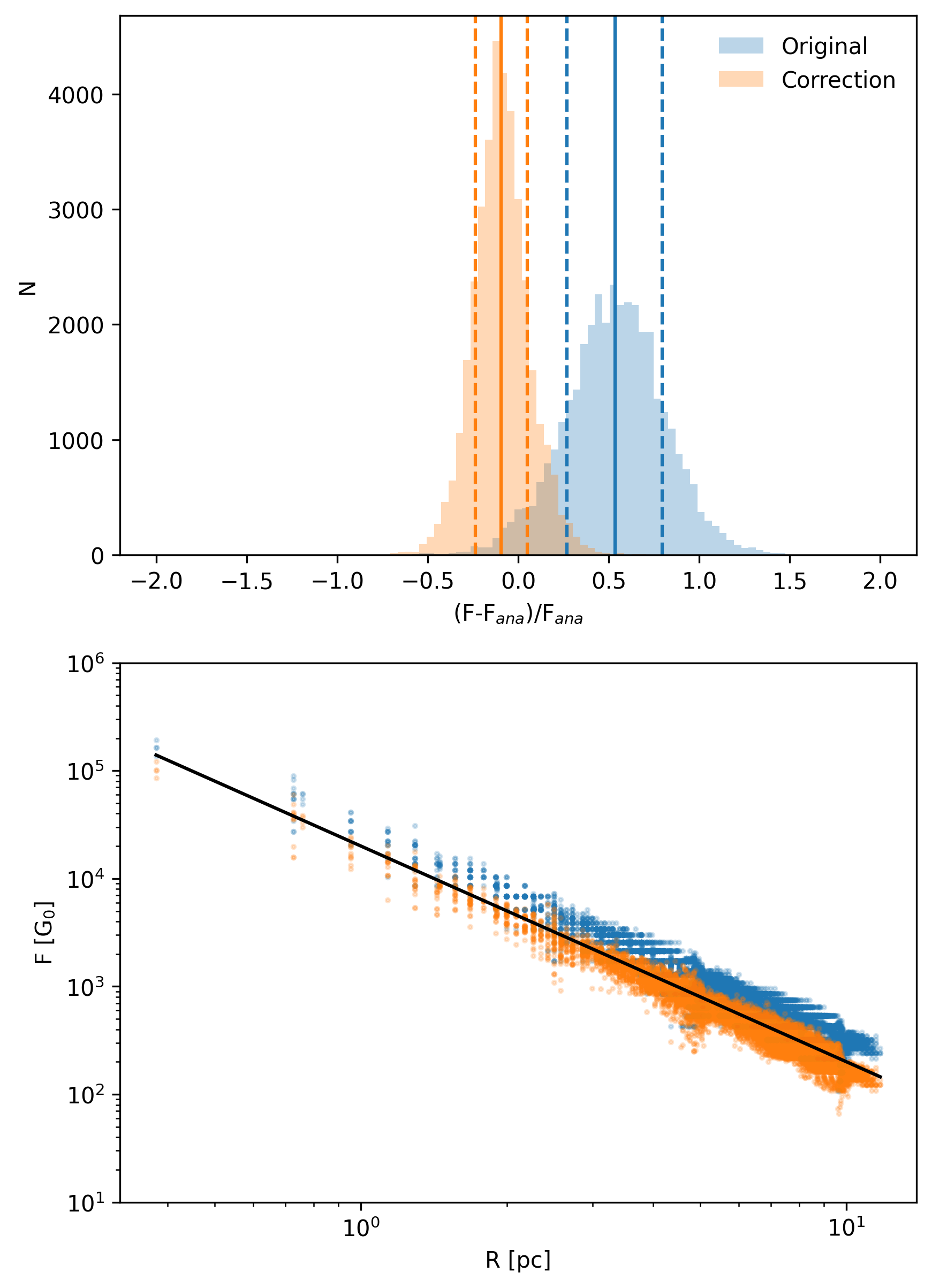}
    \caption{{\it Top panel:} The distribution of normalised differences between the analytic FUV radiation field and the grid values of a FLASH run with the original implementation (blue) and our approximate correction (orange). Solid lines denote the median, dashed lines the 16$^\textrm{th}$ and 84$^{\textrm{th}}$ percentiles. {\it Bottom panel:} The FUV radiation field as a function of radius. The analytic value is shown in black, the values in the FLASH grid cells in blue (original implementation) and orange (approximate correction).}
    \label{fig:radiation_correction_result}
\end{figure}

Under the conditions above, the exact radiation flux $F_\mathrm{ex}$ in a cell at distance $r$ from the origin is then simply $F_\textrm{ex}=\frac{L}{4\pi r^2}$. In Fig. \ref{fig:radiation_correction_result}, we show the distribution of $\Delta F = \left(F-F_\textrm{ex}\right)/F_\textrm{ex}$ (top panel) and the radiation fields in individual cells as a function of distance from the origin (bottom panel) for both methods. 

Our correction reduces the radiation field relative bias from $\sim$0.5 to $\sim$-0.1, and decreases the scatter. There is some radial structure in the deviation from the analytic case, but this appears consistent between the two implementation. 

On the whole, our approximate correction presents an improvement on the original implementation, with a smaller bias and variation, and does not appear to introduce additional artefacts. However, it remains an imperfect approximation, and an exact solution for the cross section of a cube would reduce the biased radiation field estimate even further.


\bsp	
\label{lastpage}
\end{document}